\documentclass{vldb}

\usepackage{epsfig}
\usepackage{times,graphicx,xspace,comment,subfigure,algpseudocode,multirow,url,amsmath}

\usepackage{booktabs} 
\usepackage{tabularx}
\usepackage{comment}
\usepackage{balance}  
\usepackage{enumitem}

\usepackage{algorithm}
\usepackage{algorithmicx}
\usepackage{algpseudocode}
\usepackage{setspace}
\usepackage{amssymb}
\usepackage{color}
\usepackage{flexisym}

\usepackage{balance}
\usepackage{array}
\usepackage{tabulary}
\newcolumntype{K}[1]{>{\centering\arraybackslash}p{#1}}

\usepackage{etoolbox}
\patchcmd{\maketitle}{\@copyrightpermission}{}{}{}

\DeclareSymbolFont{matha}{OML}{txmi}{m}{it}%
\DeclareMathSymbol{\q}{\mathord}{matha}{113}
\DeclareMathSymbol{\w}{\mathord}{matha}{119}

\usepackage[utf8]{inputenc}
\usepackage{amsmath}
\usepackage{amsfonts}
\usepackage{amssymb}

\usepackage{tabularx} 
\usepackage{listing}
\newcolumntype{H}{>{\setbox0=\hbox\bgroup}c<{\egroup}@{}} 

\usepackage{multirow}
\usepackage{rotating} 
\usepackage{rotate}
\usepackage{booktabs}
\usepackage{comment}
\usepackage{morefloats} 
\usepackage{nicefrac}
\usepackage{xcolor}
\usepackage{wasysym}
\newcommand{\dataName}[1]{\fontsize{6.5}{7.5}\selectfont \textsf{#1}} 
\newcommand{\eat}[1]{\ignorespaces} 
\usepackage{subfigure}

\begin{document}

\newtheorem{thm}{Theorem}
\newtheorem{theorem}{Theorem}
\newtheorem{cor}[thm]{Corollary}
\newtheorem{lemma}[thm]{Lemma}
\newcommand{\assign}{\leftarrow}
\newcommand{\be}{\begin{equation}}
\newcommand{\ee}{\end{equation}}
\newcommand{\bea}{\begin{eqnarray}}
\newcommand{\eea}{\end{eqnarray}}
\newcommand{\Mmax}{M_{\textrm{max}}}
\newcommand{\eol}{\end{enumerate}\setlength{\itemsep}{-\parsep}}
\newcommand{\eg}{\emph{e.g.}\xspace}
\newcommand{\ie}{\emph{i.e.}\xspace}
\newcommand{\etal}{\emph{et al.}\xspace}
\newcommand{\pluseq}{\mathrel{+}=}
\newcommand{\Var}{\mathrm{Var}}
\newcommand{\Cov}{\mathrm{Cov}}
\let\tilde\widetilde

\abovedisplayskip=3pt 
\abovedisplayshortskip=1pt 
\belowdisplayskip=3pt 
\belowdisplayshortskip=1pt 

\algblock{ParFor}{EndParFor}
\algnewcommand\algorithmicparfor{\textbf{parallel for}}
\algnewcommand\algorithmicpardo{\textbf{do}}
\algnewcommand\algorithmicendparfor{\textbf{end parallel}}
\algrenewtext{ParFor}[1]{\algorithmicparfor\ #1\ \algorithmicpardo}
\algrenewtext{EndParFor}{\algorithmicendparfor}

\renewcommand{\algorithmicforall}{\textbf{for each}}

\newcommand{\paraspace}{\vspace{0.05in}}
\newcommand{\parab}[1]{\paraspace\noindent{\textbf{#1}}}
\newcommand{\polylog}{\mathrm{polylog}}

\let\oldReturn\Return
\renewcommand{\Return}{\State\oldReturn}

\algnotext{EndFor}
\algnotext{EndIf}
\algnotext{EndWhile}
\algnotext{EndProcedure}
\algnotext{EndParFor}
\algnotext{EndFunction}

\def\gps{\textsc{gps}}
\def\gpsdyn{\textsc{gspsDyn}}
\def\gpsdyninc{\textsc{gspsDynInc}}

\newcommand{\Tau}{\mathrm{T}}

\let\eps\varepsilon\relax
\def\up#1{^{(#1)}}
\def\down#1{_{(#1)}}
\def\upc#1{^{\{#1\}}}
\let\hat\widehat
\def\Rl{{\mathbb R}}
\def\Nl{{\mathbb N}}
\def\Ir{{\mathbb Z}}
\def\E{{\mathbb E}}
\def\Pr{{\mathbb P}}
\def\cP{{\mathcal P}}
\def\cF{{\mathcal F}}
\def\cH{{\mathcal H}}
\def\cK{{\mathcal K}}
\def\cJ{{\mathcal J}}
\def\cZ{{\mathcal Z}}
\def\cT{{\mathcal T}}
\def\cL{{\mathcal L}}
\def\ol#1{{\overline{#1}}}
\def\var{\mathop{\mathrm{Var}}}
\def\cov{\mathop{\mathrm{Cov}}}
\def\argmin{\mathop{\mathrm{arg\,min}}}
\def\argmax{\mathop{\mathrm{arg\,max}}}
\def\nf{\hat n_{{\textsc f}}}
\def\kf{\hat K_{{\textsc f}}}
\def\kfone{\hat K_{{\textsc i}}}
\def\nfp{\hat n_{\textsc{fp}}}
\def\gsh{gSH}

\newcommand{\algrule}[1][.2pt]{\par\vskip.5\baselineskip\hrule height #1\par\vskip.5\baselineskip}

\title{On Sampling from Massive Graph Streams}

\numberofauthors{1}

\author{
\alignauthor
\begin{tabular}{K{4.1cm}K{3.7cm}K{3.5cm}K{4.5cm}}
Nesreen K. Ahmed & Nick Duffield & Theodore L. Willke & Ryan A. Rossi
\end{tabular}
\affaddr{
\begin{tabular}{K{4.1cm}K{3.7cm}K{3.5cm}K{4.5cm}}
Intel Labs & Texas A\&M University & Intel Labs & Palo Alto Research Center
\end{tabular}
}
\affaddr{
\begin{tabular}{K{4.1cm}K{3.7cm}K{3.5cm}K{4.5cm}}
nesreen.k.ahmed@intel.com &duffieldng@tamu.edu & ted.willke@intel.com &ryan.rossi@parc.com
\end{tabular}
}
}

\maketitle

\begin{abstract}
We propose Graph Priority Sampling (\gps), a new paradigm for
order--based reservoir sampling from massive streams of graph edges. \gps\
provides a general way to weight edge sampling according to auxiliary and/or size 
variables so as to accomplish various estimation goals of graph properties. In the context of subgraph counting, we show how edge sampling weights can be chosen so as to minimize the estimation variance of counts of specified sets of subgraphs. In distinction with many prior graph sampling schemes, \gps\ separates the functions 
of edge sampling and subgraph estimation. We propose two estimation frameworks: (1) Post-Stream estimation, to allow \gps\ to construct a reference sample of edges to support retrospective graph queries, and (2) In-Stream estimation, to allow \gps\ to obtain lower variance estimates by incrementally updating the subgraph count estimates during stream processing. Unbiasedness of subgraph estimators is established through a new Martingale formulation of graph stream order sampling, which shows that subgraph estimators, written as a product of constituent edge estimators are unbiased, even when computed at different points in the stream. The separation of estimation and sampling enables significant resource savings relative to previous work. We illustrate our framework with applications to triangle and wedge counting. We perform a large-scale experimental study on real-world graphs from various domains and types. \gps\ achieves high accuracy with $<1\%$ error for triangle and wedge counting, while storing a small fraction of the graph with average update times of a few microseconds per edge. Notably, for a large Twitter graph with more than 260M edges, \gps\ accurately estimates triangle counts with $<1\%$ error, while storing only 40K edges. 
\end{abstract}
\section{Introduction}
\label{sec-introduction}

The rapid growth of the Internet and the explosion in online social media has led to a data deluge. A growing set of online applications are  continuously generating data at unprecedented rates -- these range from the Internet of things (\eg, connected devices, routers), electronic communication (\eg, email, groups, IMs, SMS), social media (\eg, blogs, web pages), to the vast collection of online social networks and content sharing applications (\eg, Facebook, Twitter, Youtube, Flickr). Graphs (networks) arise as a natural data representation in many of these application domains, where the nodes represent individuals (or entities) and the edges represent the interaction, communication, or connectivity among them. 

These resulting interaction and activity networks carry a wealth of behavioral, community, and relationship information. Modeling and analyzing these massive and dynamic interaction graphs have become important in various domains. For example, detecting computer/terrorist attacks and anomalous behavior in computer networks and social media~\cite{becchetti2008efficient,aggarwal2011outlier}, identifying the behavior and interests of users in online social networks (e.g., viral marketing, online advertising)~\cite{leskovec2007dynamics,Zhaolinkpred}, monitoring and detecting virus outbreaks in human contact networks~\cite{leskovec2007cost}, among many others. But the volume and velocity of these graphs outpaces practitioners' ability to analyze and extract knowledge from them. As a result, a common practice is to analyze static windowed snapshots of these graphs over time. However, this is costly and inefficient both in terms of storage volumes, and management for future use.

To keep up with the growing pace of this data, we 
need efficient methods to analyze dynamic interaction networks as the data arrives in streams, 
rather than static snapshots of graphs. In various application domains, graph mining is rapidly shifting from mining static graph snapshots to mining an open-ended graph stream of edges representing node interactions.
We would like to have a framework capable of operating continuously and efficiently, processing edges/links as they arrive and providing timely answers for various network analysis tasks.
This motivates the streaming graph model in which the graph is presented as a stream of edges/links in any arbitrary order, where each edge can be processed only once, and any computation uses a small memory footprint (\ie, often sub-linear in the size of the input stream)~\cite{muthu,mcgregor2009graph,AhmedTKDD}. 

While studying dynamic interaction networks is important, other applications require efficient methods for the analysis of static graphs that are
too large to fit in memory~\cite{atish,muthu}. 
In these cases traditional graph methods are not appropriate as they require random disk accesses that incur large I/O costs. This naturally leads to the question: how can we process massive static graphs sequentially (one edge at a time). 
The streaming graph model would provide an ideal framework for both massive static and dynamic graphs. It would also apply to the case of graph data that is stored as a list of edges streaming from storage. Thus, any algorithm designed to process graph streams is also applicable for static graphs~\cite{AhmedTKDD}. 

Despite the recent advances in high-performance graph analysis tools and the availability of computational resources on the cloud, running brute-force graph analytics is usually too costly, too inefficient, and too slow in practical scenarios. In many cases, the cost of performing the exact computation is often not worth the extra accuracy. While an approximate answer to a query or an analysis task is usually sufficient, in particular when the approximation is performed with sufficient high-quality, unbiasedness, and confidence guarantees. 

Sampling provides an attractive approach to quickly and efficiently find an approximate answer to a query, or more generally, any analysis objective. While previous work on sampling from graph streams focused on sampling schemes for the estimation of certain graph properties (\ie, in particular triangles)~\cite{jha2013space,pavan2013counting,buriol2006counting}, in this paper however, we focus on an adaptive general purpose framework for sampling from graph streams. From a high-volume stream of edges, the proposed framework maintains a generic sample of limited size that can be used at any time to accurately estimate the total weight of arbitrary graph subsets (\ie, triangles, cliques, stars, subgraph with particular attributes). In order to obtain an accurate estimation of various graph properties, we need to maintain a \emph{weight sensitive} sample that can devote sampling resources to edges that are informative for those properties.

In addition, we want a sampling scheme that is capable of utilizing auxiliary information about the items in the stream. Most of the previous work on stream sampling is either hard to adapt to various estimation objectives, focused on specific graph properties, or incapable of utilizing auxiliary information. 

\parab{Contributions}. The main contributions of this paper are as follows. 
\begin{enumerate}[leftmargin=8pt,itemindent=10pt,itemsep=2pt,parsep=0pt,topsep=2pt,partopsep=0pt]
\item[$\bullet$] \emph{Framework.} We propose \emph{graph priority sampling} (\gps), the first adaptive, general purpose, weight sensitive, one-pass, fixed-size without replacement sampling framework for massive graph streams. \gps\ provides a general way to weight edge edge sampling according to auxiliary/size variables to estimate various graph properties (Sec~\ref{sec:framework}). We discuss antecedents to our approach in Sec~\ref{sec-known-schemes}.
\item[$\bullet$]  \emph{Theoretical Analysis.} We provide a new Martingale formulation for subgraph count estimation, and show how to compute unbiased estimates of arbitrary subgraph counts from the sample at any point during the stream; we call this \textsl{Post-Stream Estimation}, which can be used to construct reference samples for retrospective graph queries (Sec~\ref{sec:framework}).
\item[$\bullet$]  \emph{In-Stream Estimation.} We provide a second framework for \textsl{In-stream Estimation} in which subgraph count estimates are incrementally updated during stream processing rather than computed at a selected point, which can be used to obtain accurate estimates with lower variance (Sec~\ref{sec-instream}).
\item[$\bullet$]  \emph{Algorithms.} We provide efficient/parallel algorithms for triangle and wedge count estimation using \gps\ (Sec~\ref{sec:par-var}--~\ref{sec-instream}). 
\item[$\bullet$]  \emph{Accuracy.} We test our framework on graphs from various domains and types. Our estimates are accurate with $\leq 1\%$ error. Notably, for a large Twiiter graph with more than 260M edges, \gps\ obtains an accurate estimate of triangle counts with $< 1\%$ error, while storing only 40K edges (Sec~\ref{sec:eval}). 
\item[$\bullet$]  \emph{Real-time Tracking.} The proposed framework can maintain accurate real-time estimates while the stream is evolving (Sec~\ref{sec:eval}).
\end{enumerate}
We survey related work in Section~\ref{sec-related} before concluding in Section~\ref{sec:conclude}. Proofs of the Theorems are deferred to Section~\ref{sec-proofs}.

\section{Antecedents to Our Work}
\label{sec-known-schemes}

In this section, we highlight how our proposed framework generalizes many of the known sampling schemes. We discuss general statistical sampling schemes (reservoir sampling, probability proportional to size, order sampling), and highlight how graph priority sampling exploits the properties of these schemes.
 
\parab{Reservoir Sampling.} Reservoir sampling is a class of single-pass sampling schemes to sample a fixed number $n$ (possibly weighted) items from a stream of $N>n$ items \cite{Knuth:vol2,Vitter:85}.
The sample set is maintained incrementally over the stream, and can be used at any point in the stream to estimate the stream properties up to that point. In general, items have associated weight variables $w_i$ that determine non-uniform item inclusion probabilities $p_i$. 
Our graph priority sampling uses the reservoir sampling framework, to collect fixed-size weighted random sample from an input stream of edges using a single pass over the stream.

\def\pps{\textsc{PPS}}
\parab{Probability Proportional to Size Sampling.} 
In many real-world applications, an auxiliary variable (also known as a size or weight) is observed for each data item.  Auxiliary variables correlated with the population variable under study can be used as weights for non-uniform sampling.
In \emph{probability proportional to size sampling} (\pps) the inclusion probability $p_i$ is proportional to the size (or weight) $w_i$. 
Variants of this scheme have been designed to fulfill different estimation goals \cite{Tille:book}.
Inclusion Probability Proportional to Size (IPPS) is variance minimizing for a given
average sample size. It
has marginal inclusion probabilities $p_i=\min\{1,w_i/\tau\}$, so that all items of weight $w_i$ below the \textsl{threshold} $\tau$ are sampled \pps, while larger items are always sampled \cite{dlt:IEEEinfotran05}. Selected weights are assigned a Horvitz-Thompson \cite{HT52} inverse probability unbiased estimator $\hat w_i=w_i/p_i=\max\{w_i,z\}$ if $w_i$  is selected, and implicitly $0$ otherwise. 
This property inspires the sampling weights used in our edge sampling scheme, which 
minimize the incremental estimation variance in the graph properties under study.  

\parab{Order  Sampling.}
In order (or rank-based) sampling, selection depends on random order variables generated for each item, a sample of size $m$ comprising the $m$ items of highest order. The distribution of the order variables is chosen to fulfill different weighted sampling objectives, including \pps\ sampling and Weighted Sampling without Replacement \cite{Rosen1972:successive,ES:IPL2006,bottomk07:ds}.
Priority Sampling \cite{duffield2007priority} is a \pps\ scheme 
in which an item of weight $w_i$ is assigned a priority $w_i/u_i$, where the $u_i$ are IID uniform on $(0,1]$. A priority sample of size $ n$ consists of the $n$ items of highest priority, and each selected item is assigned an unbiased estimator $\hat w_i=\max\{w_i,z\}$ of $w_i$, where $z$ is the $(n+1)^{\textrm st}$ highest priority. Thus priority sampling resembles IPPS sampling with a random threshold. 

Most of the above methods are suitable mainly for sampling IID data (\eg, streams of database transactions, IP network packets). In this paper, however, we are dealing with graph data, that exhibit both structure and attributes. A few of the methods discussed above have been extended to graph streams, in particular uniform-based reservoir sampling (see in Section~\ref{sec-related}). Graph priority sampling generalizes most of the above sampling schemes, and obtains an adaptive, weight sensitive, general purpose, fixed-size sample in one-pass, while including auxiliary/size variables representing topology information that we wish to estimate.
\section{Graph Priority Sampling}
\label{sec:framework}

This section establishes a methodological
framework for graph stream priority sampling. Section~\ref{sec:prop:frame} sets up our  notation and states our estimation goals.
Section~\ref{sec:alg:desc} specifies the Graph Priority Sampling algorithm and states the properties that we will establish for it. Section~\ref{sec:unb:edge} established unbiasedness for our subgraph estimators, while Section~\ref{sec:est:var} gives unbiased estimates for the variance of these estimators. Section~\ref{sec:varopt} shows how to choose sampling weight to minimize estimation variance for target subgraphs.

\subsection{Proposed Framework}\label{sec:prop:frame}

\parab{Notation and Problem Definition}. Let $G=(V,K)$ be a graph with no self loops, where $V$ is the set of nodes, and $K$ is the set of edges. For any node $v \in V$, let $\Gamma(v)$ be the set of neighbors of node $v$ and so $\text{deg}(v) = |\Gamma(v)|$ is the degree of $v$. We call two edges $k, k' \in K$ adjacent, $k \sim\ k'$, if they join at some node, \ie, $k\;\cap\;k' \neq \emptyset$~\cite{Ahmed-gSH}. 
In this paper, we are principally concerned with estimating the frequency of occurrence of certain subgraphs of $G$. Our proposed \textsl{graph stream model} comprises an input graph $G=(V,K)$ whose edges arrive for sampling in any arbitrary order~\cite{Ahmed-gSH}. We assume edges are unique and so we can identify each edge in $K$ with its arrival order in $[|K|]=\{1,2,\dots,|K|\}$. Due to the identity between edges and arrival order we will use the notation $J \subset K$ and $J \subset [|K|]$ interchangeably. Thus, we can uniquely identify a subgraph $J \in \cJ$ with the corresponding 
ordered subset of edges $J \subset [|K|]$, written as an ordered subset, $J =  ({i_{1}},{i_{2}} ,..., {i_{\ell}})$ with $i_1 < i_2 < ... < i_{\ell}$ being the arrival order. Thus, $J \subset [t]$ if all the edges in $J$ have arrived by time $t$.

We use the general notation $\cJ$ to denote the set of all subgraphs of $G$ whose count $N(\cJ) = |\cJ|$ we wish to estimate. As special cases, $\vartriangle$ will denote the set of triangles and 
$\Lambda$ the set of wedges (paths of length $2$) in $G$,
Let $\alpha = 3 \cdot N(\vartriangle) / N(\Lambda)$ denote the global clustering coefficient of $G$. 
For a set of subgraphs $\cJ$ we shall use the notation $\cJ_t=\{J \in \cJ :  J \subset [t]\}$ to denote those members $J$ of $\cJ$ all of whose edges have arrived by time $t$,
The number of these is denoted $N_t(\cJ)=|\cJ_t|$ 

\parab{Algorithm and Intuition}. The basic outline and intuition of the proposed framework comprises of two steps. In the first step, we select a small sample of edges $\hat K_t\subset[t]$ from the set of all edges arriving by time $t$, with $m = |\hat K_t|$ is the reservoir capacity. The second step allows us to estimate the count of general subgraphs in $G$ regardless of whether they were all sampled. We define the \textbf{subset indicator} of a subset of edges  $J\subset [|K|]$ by the function, 
\be
S_{J,t} =\left\{
\begin{array}{ll}
1, & J\subset [t] \\
0, & \mbox{otherwise}\\
\end{array}
\right.
\ee
Thus, $S_{J,t}=1$ if and only if all the edges in $J$ have arrived by time $t$. In the above notation $S_{j,t}=N_t(\{j\})$ and $N_t(\cJ)=\sum_{J \in \cJ} S_{J,t}$ is the count of all members of $\cJ$ (\ie, subgraphs $J \in \cJ$) whose edges have all arrived by time $t$. 
Our goal is to estimate $N_t(\cJ)$ from a selected sample of edges $\hat K_t\subset[t]$.

\subsection{Algorithm Description \& Properties}\label{sec:alg:desc}

We formally state our main algorithm \gps($m$) for Graph Priority Sampling
into a reservoir $\hat K$ of capacity $m$ in Algorithm~\ref{alg:alg1}.
The main algorithm \gps($m$) (see Algorithm~\ref{alg:alg1}) maintains a dynamic reservoir/sample 
$\hat K$ of size $m$ from a graph whose edges are streaming over time. 
When a new edge $k$ arrives (Line~\ref{line:edge_arriv}), we call the procedure 
\textsc{$\text{GPSupdate}$}. We assume a weight function $W(k,\hat K)$ that 
expresses the sampling weight for edge $k$ as a function of both $k$ and the topology of 
the reservoir $\hat K$ (Line~\ref{line:edge_weight}). 
For example, $W(k,\hat K)$ could be set to the number of sampled edges adjacent to $k$, 
or the number of triangles in $\hat K$ completed by $k$. In general, the function $W(k,\hat K)$ 
can be set to include topology, attributes, and/or any auxiliary information in the graph. Once the weight $w(k) = W(k,\hat K)$ is computed, we assign edge $k$ a priority $r(k) = w(k)/u(k)$ (Line~\ref{line:edge_priority}), where $u(k)$ is an independent uniformly generated random number (Line~\ref{line:rnd_no}). 
\gps\ maintains a \emph{priority queue} 
where each edge in the reservoir $\hat K$ is associated with a priority (computed at arrival time) that defines its position in the heap. When a new edge $k$ arrives in the stream (and if the reservoir is full, see Lines~\ref{line:heap_up_start}--\ref{line:heap_up_end}), its priority is computed and compared with the lowest priority edge in the queue. If edge $k$ has a lower priority, then it is discarded.  If edge $k$ has a higher priority, then the lowest priority edge is discarded and replaced by edge $k$. 

\parab{Implementation and data structure}. We implement the priority
queue as a \emph{min-heap}~\cite{Cormen:2001}, where each edge has a
priority less than or equal to its children in the heap, and the root
position points to the edge with the lowest priority. Thus, access to
the lowest priority edge is performed in $O(1)$. If edge $k$ has a
higher priority than the root, edge $k$ is initially inserted at the
root position, then moved downward to its correct position in the heap
in $O (\log m)$ time (worst case). Note that if the sample size is
less than the reservoir capacity, \ie, $|\hat K| < m$, edge $k$ is
inserted in the next available position in the heap, then moved upward
to its correct position in the heap in $O (\log m)$ time (worst
case). 
The threshold $z^*$ is the $(m+1)^{\text{st}}$ highest priority (see
Line~\ref{line:z_update}). To simplify the analysis, we provisionally admit a new edge $k$ to the reservoir, then one of the $m+1$ edges is discarded if it has the lowest priority. Finally, at any time in the stream, we can call the procedure \textsc{$\text{GPSnormalize}$} to obtain the edge sampling probabilities.
As shown in the proof of Theorem~\ref{thm:single},  
$p(k') = \text{min}\{1, w(k')/z^*\}$ (see
Lines~\ref{line:prob_start}--\ref{line:prob_end}) is the conditional
sampling probability for $k'$ given $z^*$; hence $1/p(k')$ forms the Horvitz-Thompson estimator for the indicator of $k'$.
The proposed framework \gps\ naturally leads to a family of sampling algorithms 
that can be tuned for various graph sampling/analysis objectives. For example, if we set 
$W(k,\hat K) = 1$ for every $k$, Algorithm~\ref{alg:alg1} leads to uniform sampling as in the standard reservoir sampling (see~\cite{Vitter:85}). 

{
\algrenewcommand{\alglinenumber}[1]{\fontsize{6.5}{7}\selectfont#1}
\algnotext{endpar}
\begin{figure}[t!]
\vspace{-3mm}
\begin{center}
\begin{minipage}{1.0\linewidth}
\begin{algorithm}[H]
\caption{\;
Family of Graph Priority Sampling Algorithms
}
\label{alg:alg1}
\begin{spacing}{1.1}
\fontsize{8}{9}\selectfont
\algrenewcommand{\alglinenumber}[1]{\fontsize{6.5}{7}\selectfont#1}
\begin{algorithmic}[1]
\Procedure{\textsc{$\text{GPS}$}}{$m$}
\State$\hat K \assign\emptyset$; $z^*\assign 0$ 
\While{new edge $k$} \label{line:edge_arriv}
  \State \Call{\textsc{$\text{GPSupdate}$}}{$k, m$}
\EndWhile
 \State \Call{\textsc{$\text{GPSnormalize}$}}{$\hat K$}
\EndProcedure
\algrule
\Procedure{\textsc{$\text{GPSupdate}$}}{$k, m$}
  \State Generate $u(k) \sim\ Uni(0,1]$ \label{line:rnd_no}
  \State $w(k)\assign W(k,\hat K)$ \label{line:edge_weight}
  \State $r(k)\assign w(k)/u(k)$ \Comment{Priority of edge $k$} \label{line:edge_priority}
  \State $\hat K\assign \hat K\cup\{k\}$ \Comment{Provisionally include edge $k$} \label{line:add_item_prov}
  \If{$|\hat K| > m$} \label{line:heap_up_start}
  \State $k^*\assign\argmin_{k'\in \hat K}r(k')$ \Comment{Lowest priority edge}
  \State $z^*\assign\max\{z^*,r(k^*)\}$ \label{line:z_update}\Comment{New threshold}
  \State $\hat K \assign \hat K\setminus\{k^*\}$ \Comment{Remove lowest priority edge}  \label{line:heap_up_end}
  \EndIf
\EndProcedure
\algrule
\Procedure{\textsc{$\text{GPSnormalize}$}}{$\hat K$}
 \For{$k'\in \hat K$} \label{line:prob_start}
 \State $p(k')\assign \min\{1,w(k')/z^*\}$ \Comment{HT Renormalize} \label{line:prob_end}
 \EndFor
\EndProcedure
\smallskip
\end{algorithmic}
\end{spacing}
\end{algorithm}
\end{minipage}
\vspace{-6mm}
\end{center}
\end{figure}
}

\parab{Algorithm Properties}. Graph Priority Sampling demonstrates the following properties:
\parab{(S1) Fixed Size Sample.} As seen above, $\hat K_t$ is a 
reservoir sample of fixed size $|\hat K_t| =m$ for all $t\ge m$.

\parab{(S2) Unbiased Subgraph Estimation.}
In Section~\ref{sec:unb:edge} we construct unbiased subgraph
estimators $\hat S_{J,t}$ of $S_{J,t}$ for each subgraph $J$ and
$t>0$. The $S_{J,t}$ is computable from the sample sets $\hat K_t$. 
Section~\ref{sec-instream} extends our construction to new classes of
estimators $\prod_{i\in J} \hat S_{i,t_i}$ that are edge products over multiple
times. These allow unbiased estimation of subgraphs in new ways: as they arise
in the sample, or prior to discard, or on arrival of certain
edges. These results follow from a novel Martingale formulation of graph priority sampling.

\parab{(S3) Weighted Sampling and Variance Optimization.}
Graph priority sampling provides a mechanism to tune sampling weights to the needs of
applications. We accomplish this using edge weights that
express the role of an edge in the sampled graph. Examples 
include the number of edges in the currently sampled graph
that are adjacent to an arriving edge, and the number of subgraphs
bearing a given motif that would be created by inclusion of the edge
in the sample.
Section~\ref{sec:varopt} shows how to choose weights to minimize the 
variance of estimated counts of specific target subgraphs. Weights may also express \textsl{intrinsic} properties that do not depend explicitly on the graph structure. Examples include endpoint node/edge identities, attributes, and other auxiliary variables, such as user age, gender, interests, or relationship types in social networks, and bytes associated with communication links in technological/IP networks.

\parab{(S4) Computational Feasibility.} For each arriving edge $k = (v_1,v_2)$, the \gps\ framework calls \textsc{$\text{GPSupdate}$} to update reservoir $\hat K$ of capacity $m$. The processing time for a new arrival comprises of the cost to compute the weight (\ie, $W(k, \hat K)$ and the cost to update the heap (if the new edge is inserted). We use a binary heap implemented by storing the edges in a standard array and using their relative positions within that array to represent heap (parent-child) relationships. 
The worst case cost of binary heap insertion/deletion is $O(\log
m)$. The cost of $W(k, \hat K)$ is problem-specific and depends on the
sampling objective and the function that computes the sampling weight
for edge $k$. 
We use the number of triangles in $\hat K$ completed by $k$, \ie, $W(k, \hat K) = | \hat \Gamma(v_1) \cap \hat \Gamma(v_2)|$. This can be achieved in $O(\text{min}\{\text{deg}(v_1),\text{deg}(v_2)\})$, if a hash table or a bloom filter is used for storing $\hat \Gamma(v_1), \hat \Gamma(v_2)$ and looping over the sampled neighborhood of the vertex with minimum degree and querying the hash table of the other vertex. 
The space requirements of \gps\ is: $O(|\hat V| + m)$, where $|\hat V|$ is the number of nodes in the reservoir, and $m$ is the reservoir capacity. In general, there is a trade-off between space and time, and \gps\ could limit the space to $O(m)$, however, the cost update per edge would require a pass over the reservoir ($O(m)$ in the worst case). On the other hand, if we increase the space to $O(|\hat V| + m)$, then we can achieve a sub-linear time for edge updates.

\subsection{A Martingale for Subgraph Counting}\label{sec:unb:edge}
We now axiomatize the dependence of $w$ on $k$ and $\hat K$ and
analyze the statistical properties of estimates based on the sample
set. We index edge arrivals by $t\in\Nl$, and let $\hat K_t\subset[t]$
denote the set of indices in the sample after arrival $t$ has been
processed and $\hat K'_t=\hat K_{t-1}\cup\{t\}$ denote the index set after $t$
has been provisionally included in the sample. Let $w_t$ denote the weight assigned to arrival $t$ and
$u_t$ be IID uniform on $(0,1]$. The priority of arrival $t$ is then
$r_t=w_t/u_t$. An edge $i\in\hat K'_t$ is
selected if it does not have the smallest priority in $\hat K'_t$,
i.e., if
\be r_i > z_{i,t} = \min_{j\in\hat K'_t\setminus\{i\}}r_j
\ee
When $i$ is selected from $\hat K'_t$ then $r_{i,t}$ is equal to the unrestricted
minimum priority $z_{t}=\max_{j\in  \hat K'_{t}}r_j$ since the
discarded edge takes the minimum. For $t<i$, $z_{i,t}=z_t$ since $i$ has not yet appeared.
Defining $B_{i}(x) = \left\{r_{i} > x \right\}$, we write the event that $i$ is
in the sample at time $t\ge i$ as
\be
\{ i\in \hat K_t\} = \cap_{i=s}^t B_{i}(z_{i,s})
\ee

We now construct for each edge $i$ a sequence of \textbf{Edge Estimators}
$\hat S_{i,t}$ that are unbiased estimators of the corresponding edge
indicators. 
We prove unbiasedness by establishing that the sequence is a Martingale
~\cite{W91}. A sequence of random variables $\{X_t :
t\in\Nl\}$ is a \textsl{Martingale} with respect to a filtration $\cF=\{\cF_t:
t\in\Nl\}$ of increasing $\sigma$-algebra (these can be regarded as
variables for conditioning) if each $X_t$ is
measurable w.r.t. $\cF_t$ (i.e. it is function of the
the corresponding variables) and obeys:
\be
\E[X_{t}|\cF_{t-1}]=X_{t-1}
\ee
Martingales provide a framework within which to express
unbiased estimation in nested sequences of random variables.

We express \gps\ within this framework. For $J\subset \Nl$ let 
$z_{J,t}=\min_{j\in\hat K'_t\setminus J}r_j$. Let 
$\cF\up 0_{i,t}$ denote the $\sigma$-algebra generated by the
variables $\{B_j(z_{\{ij\},s}): j\ne i,\ s\le t \}$, let
$\cF_{i,t}$ be the $\sigma$-algebra generated by $\cF\up0_{i,t}$
and the variables $\cZ_{i,t} = \{z_{i,s}: s\le t\}$ and
$\{B_{i}(z_{i,t}),i\le s\le t\}$, and let $\cF_{i}$ be
the filtration
$\{\cF_{i,t}:t\ge i-1\}$.

Set
$z^*_{i,t}=\max_{i\le s \le t}z_{i,s}$ and define
\be\label{eq:mart2}
R_{i,t}= \min\{1,w_i/z^*_{i,t}\}
\mbox{ and }
\hat S_{i,t}
=\frac{I(B_{i}(z^*_{i,t}))}{R_{i,t}}
\ee
for $t\ge i$ and $\hat S_{i,t}=0$ for $0\le t<i$.
\begin{theorem}[Edge Estimation]\label{thm:single}
Assume $w_{i,t}$ is $\cF_{i,t-1}$-measurable. Then
$\{\hat S_{i,t}: t\ge i\}$ is a Martingale w.r.t. the
filtration $\cF_i$, and hence  $\E[\hat S_{i,t}]=S_{i,t}$ for all $t\ge 0$
\end{theorem}
The measurability condition on $w_i$ means that it is determined by
the previous arrivals, including the case that an edge weight depends
on the sampled topology that it encounters on arrival.

To compute $\hat S_{i}$ we observe that when $i\in\hat K_t$, then
(i) $r_i>z_{i,t}$ and hence $z_{i,t}=z_t$; and (ii)
$z_{i,i}>\max_{s\le i} z_{i,s}$ since otherwise the minimizing $j$ for
$z_{i,i}$ could not have been selected. Hence for $t\ge i$ with
$z^*_t=\max_{s\le t}z_s$:
\be
\hat S_{i,t} 
=\frac{I(i\in\hat K_t)}{\min\{1,w_i/z^*_{t}\}}
\ee

It is attractive to posit the product $\hat S_J=\prod_{i\in J}\hat
S_i$ as a subgraph estimator $J$ when $J\subset [t]$. While this
estimator would be unbiased for independent sampling, the constraint
of fixed-size introduces dependence between samples and hence bias of
the product.  For example, VarOpt sampling 
\cite{Cohen:2011:ESS:2079108.2079117} obeys only $\E[\prod_{i\in J}\hat S_i]\le \prod_{i\in
  J}\E[\hat S_i]$. We now show that the edge estimators for Graph
Priority Sampling, while dependent, have zero correlation. This is a consequence of the property, that we now establish, that the edge product estimator is a Martingale.

Fix a subgraph as $J\subset\Nl$, set $J_t=J\cap[t]$ and for $k\in[t] \cap J^c$ 
define $z_{J,k,t}=\max_{i\in \hat K'_t\setminus
    (J_t\cup\{k\})}r_{i}$, i.e., the maximal rank in $\hat K'_t$ apart
  from $k$ and those in $J_t$. Let $\cF\up 0_{J,t}$ denote the
  conditioning w.r.t $\{B_k(z_{J,k,s}):k\notin J, s\le t\}$, and let
  $\cF_{J,t}$ denote conditioning w.r.t $\cF\up 0_{J,s}$,
  $\cZ_{J,t}=\{z_{J,s}:s\le t\}$ and $\{B_i(z_{J,s}): i\in J, s\le
  t\}$ and let $\cF_J$ denote the corresponding filtration.

\begin{theorem}[Subgraph Estimation]\label{thm:prod}
\begin{itemize}
\item[(i)] For $J\subset \Nl$ define $S_{J,t}=\prod_{i\in J}\hat S_{i,t}$. Then $\{\hat S_{J,t}: t\ge \max J\}$ is a
Martingale w.r.t. the filtration $\cF_J$ and hence $\E[\prod_{i\in
  J}\hat S_{i,t}]=S_{J,t}$ for $t\ge 0$.
\item[(ii)] For any set $\cJ$ of subgraphs of $G$, $
\hat N_t(\cJ)=\sum_{J\in \cJ: J\subset K_t}\hat S_{J,t}$
is an unbiased estimator of $N_t(\cJ) =|\cJ_t|=\sum_{J\in\cJ}S_{J,t}$, and the sum can be restricted to those $J\in \cJ$ for which $J\subset\hat K_t$, i.e., entirely within the sample set at $t$.
\end{itemize}
\end{theorem}

The proof of Theorem~\ref{thm:prod}(i) mirrors that of
Theorem~\ref{thm:single}, and follows from the fact that
the expectation of $\prod_{j\in J}B_j(z_{J,t})$, conditional on
  $\cF_{J,t-1}$, is a product over $J$; we omit the details. Part (ii) follows by linearity of 
expectation, and the restriction of the sum follows since $\hat S_{J,t}=0$ unless $J\subset \hat K_t$.

\subsection{Variance and Covariance Estimation}\label{sec:est:var}

Theorem~\ref{thm:prod} also allows us to establish unbiased estimators
for the variance and covariance amongst subgraph estimators. Consider two edge subsets $J_1, J_2 \subset \hat K_t$. We use the following as an estimator of $\cov(\hat S_{J_1,t},\hat S_{J_2,t})$: 
\begin{align}\label{eq:cov:est:post}
\hat C_{J_1,J_2,t} &= \hat S_{J_1,t} \hat S_{J_2,t} - \hat S_{J_1\setminus J_2,t} \hat S_{J_2\setminus J_1,t}\hat S_{J_1\cap J_2,t}  \nonumber \\
&= \hat S_{J_1\setminus J_2,t} \hat S_{J_2\setminus J_1,t}\hat S_{J_1\cap J_2,t} \bigg(\hat S_{J_1 \cap J_2,t} - 1 \bigg) \nonumber \\
&= \hat S_{J_1 \cup J_2,t} \bigg(\hat S_{J_1 \cap J_2,t} - 1\bigg)
\end{align}
\begin{theorem}[Covariance Estimation]
\label{thm:post:cov}
$\hat C_{J_1,J_2,t}$ is an estimator of $\cov(\hat S_{J_1,t},\hat S_{J_2,t})$.
\begin{enumerate}[label=(\roman*)]
\item $\hat C_{J_1,J_2,t}$ is an unbiased
  estimator of $\cov( \hat S_{J_1,t} , \hat S_{J_2,t})$.
\item $\hat C_{J_1,J_2,t}\ge0$ and hence
  $\cov( \hat S_{J_1,t} , \hat S_{J_2,t})\ge0$.
\item $\hat S_{J,t}(\hat S_{J,t} -1)$ is an unbiased
  estimator of $\var(\hat S_{J,t})$.
\item $\hat C_{J_1,J_2,t}=0$ if and only if $\hat
  S_{J_1,t}=0$ or $\hat
  S_{J_2,t}=0$, or $J_1\cap J_2 = \emptyset$, i.e., covariance estimators are computed only
  from edge sets that have been sampled and their intersection is non-empty.
\end{enumerate}
\end{theorem}
We do not provide the proof since these results are a special case of a more general result that we
establish in Section~\ref{sec-instream} product form graph estimators
in which the edges are sampled at different times.

\subsection{Optimizing Subgraph Variance}\label{sec:varopt}

How should the ranks $r_{i,t}$ be distributed in order to minimize the
variance of the unbiased estimator $\hat N_t(\cJ)$ in Theorem~\ref{thm:prod} ?
This is difficult to formulate directly because
the variances of the $\hat S_{j,t}$ cannot be computed simply from the candidate
edges. Instead, we minimize the conditional variance of the increment in $N_t(\cJ)$ 
incurred by admitting the new edge to the sample:

To be precise:
\begin{enumerate}[leftmargin=0pt,itemindent=10pt,itemsep=0pt] 
\item
For each arriving edge $i$ find the marginal selection probability for $i$
that minimizes the conditional variance $\var(\hat
N_{i}(\cJ)|\cF_{i,i-1})$. 
\item
Edges are priority order sampled using weights that implement the variance-minimizing selection 
probabilities.
\end{enumerate}
Our approach is inspired by the cost
minimization approach of IPPS sampling \cite{dlt:IEEEinfotran05}. When $i\in\hat K'_t$ we define $\hat N_{i,t}(\cJ) = \#\{J \in \cJ : i \in J \wedge J \subset \hat K'_t\}$
i.e., the number of members $J$ of $\cJ$ that are subsets of the
set of candidate edges $\hat K'_t$ and that contain $i$. Put another
way, $\hat N_{i,t}(\cJ)$ is the number of members of $\cJ$ that
are created by including $i$ within $\hat K_{t}$. Suppose $i$ is sampled
with probability $p$, conditional on $\cF_{i,i-1}$.  The expected space
cost of the sampled is proportional to $p$, while the sampling variance
associated with Horvitz-Thompson estimation of the increment
$n=\hat N_{i,t}(\cJ)$ is $ n^2(1/p-1)$. Following \cite{dlt:IEEEinfotran05},
we form a composite cost
\be
C(z) = z^2 p + n^2(1/p-1)
\ee
where $z$ is a coefficient expressing the relative scaling of the
two components in the cost. $C(z)$ is minimized by
$p=\min\{1,n/z\}$, corresponding to IPPS sampling with
threshold $z$ and weight $n$. By comparison with the
relation between threshold sampling and priority sampling, this
suggests using $n=\hat N_{i,t}(\cJ)$ as the weight for graph stream order
sampling.
We also add a default weight for each edge so that an arriving edge $k$ that does not currently intersect with the target class (i.e. $k\ne J\subset \hat N_t(\cJ)$) can be sampled.

\section{Triangle \& Wedge Counting}
\label{sec:par-var}

In this section we apply the framework of Section~\ref{sec:framework} to triangle and wedge counting, and detail the computations involved for the unbiased subgraph count estimates and their variance estimates.

\parab{Unbiased Estimation of Triangle/Wedge Counts.} From the notation in Sec.~\ref{sec:framework}, let $\vartriangle_t$ denote the set of triangles whose all edges have arrived by time $t$, and $\hat \vartriangle_t \subset \vartriangle_t$ be the subset of such triangles that appear in the sample $\hat K_t$. Then, $\hat N_{t} (\vartriangle)$ is the Horvitz-Thompson estimator of the count of members (\ie, triangles) in $\vartriangle_t$. We write $\tau \in \vartriangle_t$ as a subset $(k_1, k_2, k_3)$ ordered by edge arrival (\ie, $k_3$ is the last edge). Similarly, $\Lambda_t$  denote the set of wedges whose all edges have arrived by time $t$. So $\hat N_{t} (\Lambda)$ is the Horvitz-Thompson estimator of the count of wedges in $\Lambda_t$, and  $\lambda \in \Lambda_t$ is written as an ordered subset $(k_1,k_2)$ with $k_2$ the last edge.  The following are direct corollaries of Theorem~\ref{thm:prod}:
\begin{cor}[Triangle Count]
\label{thm:post:HT_trian}
$\hat N_{t} (\vartriangle) = \sum_{\tau \in \vartriangle_t} \hat S_{{\tau},t}$ is an unbiased estimator of $N_{t} (\vartriangle)$. 
\end{cor}
\begin{cor}[Wedge Count]
\label{thm:post:HT_wedge}
$\hat N_{t} (\Lambda) = \sum_{\lambda \in \Lambda_t} \hat S_{{\lambda},t}$ is an unbiased estimator of $N_{t} (\Lambda)$. 
\end{cor}
\vspace*{-0.25cm}

Additionally, we use $\hat \alpha_t = 3 \hat N_{t} (\vartriangle) / \hat N_{t} (\Lambda)$ as an estimator for the global clustering coefficient $\alpha_t$. 

\parab{Variance Estimation of Triangle/Wedge Counts.} Let 
$\var[\hat N_t(\vartriangle)]$ denote the variance of the unbiased estimator of triangle count at time $t$, and $\var[\hat N_t(\Lambda)]$ the variance of the unbiased estimator of wedge count at time $t$, given by Corollaries~\ref{thm:post:HT_trian} and  \ref{thm:post:HT_wedge} respectively. 
Expressions for unbiased estimates of these variances are direct corollaries from Theorem~\ref{thm:post:cov}, which itself follows from Theorem~\ref{thm:snap:cov}.
\vspace*{-1mm}
\begin{cor}[Variance of Triangle Count]
\label{thm:post:var_tri}
$\hat V_{t} (\vartriangle) $ is an unbiased estimator of $\mathrm{Var}[\hat N_t(\vartriangle)]$, where
\end{cor}
\vspace*{-2mm}
\begin{align}
\hat V_{t} (\vartriangle) = \sum\limits_{\tau \in \vartriangle_t} \hat S_{{\tau},t} ( \hat S_{{\tau},t} -1) + 2 \sum\limits_{\tau \in \vartriangle_t} \sum\limits_{\substack{\tau' < \tau \\ \tau' \in \vartriangle_t}} \hat C_{\tau,\tau',t} \label{eq:var:est:tri}
\end{align}
\begin{cor}[Variance of Wedge Count]
\label{thm:post:var_wedge}
$\hat V_{t} (\Lambda)$ is an unbiased estimator of $\mathrm{Var}[\hat N_t(\Lambda)]$, where
\end{cor}
\vspace*{-3mm}
\begin{align}
\hat V_{t} (\Lambda) = \sum\limits_{\lambda \in \Lambda_t} \hat S_{{\lambda},t} ( \hat S_{{\lambda},t} -1) + 2 \sum\limits_{\lambda \in \Lambda_t} \sum\limits_{\substack{\lambda' < \lambda \\ \lambda' \in \Lambda_t}} \hat C_{\lambda,\lambda',t} \label{eq:var:est:wedge}
\end{align}

\parab{Variance Estimation for Global Clustering Coefficient.}
We use $\hat \alpha = 3\hat N_t(\vartriangle) /\hat N_t(\Lambda)$ as an
estimate of the global clustering coefficient $\alpha=3N(\vartriangle)/N(\Lambda)$.
While this estimate is biased, asymptotic convergence to the true value for large graphs would follow form the property for  $\hat N_t(\vartriangle)$ and $\hat N_t(\Lambda)$.
This motivates using a Taylor expansion of the estimator, using the asymptotic form of the well-known delta-method \cite{schervish} in order to approximate its variance; see 
\cite{Ahmed-gSH} for a similar approach for Graph Sample and Hold. The resulting approximation is:
\vspace*{-2mm}
\bea\label{eq:var:cluster} \kern -10pt
\var (\hat N(\vartriangle)/\hat N(\Lambda))\kern-5pt &\approx& \kern -5pt
\frac{\var(\hat N(\vartriangle))}{\hat N(\Lambda)^2}+\frac{\hat N(\vartriangle)^2 \var(\hat N(\Lambda) )}{\hat N(\Lambda)^4} \\
&&-2\frac{\hat N(\vartriangle)\cov(\hat N(\vartriangle),\hat N(\Lambda))}{\hat N(\Lambda)^3} \nonumber
\eea
Following
Theorem~\ref{thm:post:cov}, the covariance $\cov(\hat N(\vartriangle),\hat N(\Lambda))$ is
estimated as
\vspace*{-2mm}
\be \hat V(\vartriangle,\Lambda)=
\sum_{\tau\in\hat\vartriangle,\lambda\in\hat\Lambda \atop \tau\cap \lambda \ne\emptyset}
\hat S_{\tau\cup \lambda}\left(\hat S_{\tau\cap \lambda}-1\right)
\ee

\parab{Efficiency.} The basic intuition of Algorithm~\ref{alg:alg_gpsest} is that the subgraph estimation problem is localized. Hence, all computations can be efficiently performed by exploring the \emph{local neighborhood} of an edge (or a node)~\cite{ahmed2015efficient}. In this section, we discuss how the estimators can be adapted to make the computations more efficient and localized, while still remaining unbiased.

By linearity of expectation, we express the unbiased estimator $\hat N_{t} (\vartriangle)$ (from Theorem~\ref{thm:post:HT_trian}) as $\hat N_{t} (\vartriangle) = 1/3 \sum_{k \in \hat K_t} \hat N_{k,t} (\vartriangle)$ where $\hat N_{k,t} (\vartriangle)$ is the conditional estimator of triangle counts for edge $k$ normalized by the number of edges in a triangle. Similarly, we express the unbiased estimator $\hat N_{t} (\Lambda)$ (from Theorem~\ref{thm:post:HT_wedge}) as $\hat N_{t} (\Lambda) = 1/2 \sum_{k \in \hat K_t} \hat N_{k,t} (\Lambda)$ where $\hat N_{k,t} (\Lambda)$ is the conditional estimator of wedge counts for edge $k$ normalized by the number of edges in a wedge. 

Consider any two distinct edge subsets $J_1, J_2 \subset \hat K_t$. From Theorem~\ref{thm:post:cov}, the covariance estimator $\hat C_{J_1,J_2,t} = 0$ if $J_1$ and $J_2$ are disjoint (\ie, $|J_1 \cap J_2| = 0$). Otherwise, $\hat C_{J_1,J_2,t} > 0$ if and only if their intersection is non-empty (\ie, $|J_1 \cap J_2| > 1$). If $J_1, J_2$ are triangles (or wedges), then $|J_1 \cap J_2| \leq 1$, since any two \emph{distinct} triangles (or wedges) could never overlap in more than one edge. Thus, the unbiased variance estimators can also be computed locally for each edge, as we show next. 

By linearity of expectation, we re-write Equation~\ref{eq:var:est:tri} as follows:

\bea
\hat V_{t} (\vartriangle) &=&  1/3 \sum\limits_{k \in \hat K_t}  \sum\limits_{\tau \in \vartriangle_t(k)} \hat S_{{\tau},t} ( \hat S_{{\tau},t} -1) \\ && \kern -40pt + \sum\limits_{k \in \hat K_t}  \sum\limits_{\tau \in \vartriangle_t(k)} \sum\limits_{\substack{\tau' < \tau \\ \tau' \in \vartriangle_t(k)}} \hat S_{\tau \setminus \tau',t} \hat S_{\tau' \setminus \tau,t}\hat S_{\tau \cap \tau',t} \bigg(\hat S_{\tau \cap \tau',t} - 1 \bigg) \nonumber
\eea

Note that for any two distinct triangles $\tau, \tau' \subset \hat K_t$, we have $\hat S_{\tau \cap \tau', t} > 0$ if and only if $\tau \cap \tau' = \{k\}$ for some edge $k \in \hat K_t$. Similarly, we could re-write Equation~\ref{eq:var:est:wedge} as follows:

\bea
\hat V_{t} (\Lambda) &=&  1/2 \sum\limits_{k \in \hat K_t}  \sum\limits_{\lambda \in \Lambda(k)} \hat S_{{\lambda},t} ( \hat S_{{\lambda},t} -1) \\ && \kern -40pt + \sum\limits_{k \in \hat K_t}  \sum\limits_{\lambda \in \Lambda_t(k)} \sum\limits_{\substack{\lambda' < \lambda \\ \lambda' \in \Lambda_t(k)}} \hat S_{\lambda \setminus \lambda',t} \hat S_{\lambda' \setminus \lambda,t}\hat S_{\lambda \cap \lambda',t} \bigg(\hat S_{\lambda \cap \lambda',t} - 1 \bigg)
\nonumber \eea

\parab{Algorithm Description.} To simplify the presentation of Algorithm~\ref{alg:alg_gpsest}, we drop the variable $t$ that denote the arrival time in the stream, however the algorithm is valid for any $t \in \Nl$. We start by calling Algorithm~\ref{alg:alg1} to collect a \emph{weighted sample} $\hat K$ of capacity $m$ edges. For each edge $k \in \hat K$, we use $W(k,\hat K) = 9 * |\hat \vartriangle(k)| + 1$ where $|\hat \vartriangle(k)|$ is the number of triangles completed by edge $k$ and whose edges in $\hat K$. Then, we call Algorithm~\ref{alg:alg_gpsest} at any time $t$ in the stream to obtain unbiased estimates of triangle/wedge counts, global clustering, and their unbiased variance. 

For each edge $k=(v_1, v_2) \in \hat K$, Alg.~\ref{alg:alg_gpsest} searches the neighborhood $\hat \Gamma(v_1)$ of the node with minimum degree (\ie, $\text{deg}(v_1) \leq \text{deg}(v_2)$ for triangles (Line~\ref{alg2:search_min_deg}). Lines~\ref{alg2:tri:start}--\ref{alg2:tri:end} compute the estimates for each triangle $(k_1, k_2, k)$ incident to edge $k$. Lines~\ref{alg2:wedg1:start}--\ref{alg2:wedg1:end} compute the estimates for each wedge $(k_1, k)$ incident to edge $k$ (and centered on node $v_1$). Then, Lines~\ref{alg2:wedg2:start}--\ref{alg2:wedg2:end} compute the estimates for each wedge $(k_2, k)$ incident to edge $k$ (and centered on node $v_2$). Finally, the individual edge estimators are summed in  Lines~\ref{alg2:reduce:start}--\ref{alg2:reduce:end}, and returned in Line~\ref{alg2:return}.

We state two key observations in order: First, the estimators of triangle/wedge counts can be computed locally for each sampled edge $k \in \hat K$, while still remaining unbiased. Thus, Algorithm~\ref{alg:alg_gpsest} is localized. Second, since the estimators of triangle/wedge counts can be computed for each sampled edge $k \in \hat K$ independently in parallel, Algorithm~\ref{alg:alg_gpsest} already has abundant parallelism.

\parab{Complexity.} Algorithm~\ref{alg:alg_gpsest} has a total runtime of $O(m^{3/2})$. This is achieved by $\sum_{(v_1,v_2) \in \hat K}{\text{min}\{\text{deg}(v_1),\text{deg}(v_2)\})} = O(a(\hat K) m)$, where $a(\hat K)$ is the arboricity of the reservoir graph. This complexity can be \emph{tightly bounded} by $O(m^{3/2})$ since $O(a(\hat K) m) \leq O(m^{3/2})$ for any sampled graph~\cite{chiba1985arboricity,ahmed2015efficient}. 

{
\algrenewcommand{\alglinenumber}[1]{\fontsize{6.5}{7}\selectfont#1}
\algnotext{endpar}
\begin{figure}[t!]
\begin{center}
\begin{minipage}{1.0\linewidth}
\begin{algorithm}[H]
\caption{\;
Unbiased Estimation of Triangle \& Wedge Counts
}
\label{alg:alg_gpsest}
\begin{spacing}{1.1}
\fontsize{8}{9}\selectfont
\algrenewcommand{\alglinenumber}[1]{\fontsize{6.5}{7}\selectfont#1}
\begin{algorithmic}[1]
\Procedure{\textsc{$\text{GPSestimate}$}}{$\hat K$}
\State Initialize all variables to zero
\ParFor{edge $k=(v_1,v_2) \in \hat K$}
\State $q \assign \text{min}\{1,w(k)/z^*\}$
\ForAll{$v_3 \in \hat \Gamma(v_1)$} \Comment{found wedge} \label{alg2:search_min_deg}
\State $k_1 \assign (v_1,v_3)$
\State $q_1 \assign \text{min}\{1,w(k_1)/z^*\}$
\vspace{0.5mm}
\State /*Compute triangle estimates*/
\If {$v_3 \in \hat \Gamma(v_2)$}  \Comment{found triangle} \label{alg2:tri:start}
\State $k_2 \assign (v_2,v_3)$
\State $q_2 \assign \text{min}\{1,\w(k_2)/z^*\}$
\State $\hat N_k(\vartriangle) \pluseq (q q_1 q_2)^{-1}$  \Comment{triangle count}
\State $\hat V_k(\vartriangle) \pluseq (q q_1 q_2)^{-1} ((q q_1 q_2)^{-1} -1)$ \Comment{tri. var.}
\State $\hat C_k(\vartriangle) \pluseq c_{\vartriangle} * (q_1 q_2)^{-1}$ \Comment{triangle covariance}
\State $c_{\vartriangle} = c_{\vartriangle} + (q_1 q_2)^{-1}$ 
\EndIf \label{alg2:tri:end}
\vspace{0.5mm}
\State /*Compute wedge estimates for wedges $(v_3,v_1,v_2)$*/ 
\State $\hat N_k(\Lambda) \pluseq (q q_1)^{-1}$ \Comment{wedge count} \label{alg2:wedg1:start}
\State $\hat V_k(\Lambda) \pluseq (q q_1)^{-1} ((q q_1)^{-1} -1)$ \Comment{wedge variance}
\State $\hat C_k(\Lambda) \pluseq c_{\Lambda} * q_1^{-1}$ \Comment{wedge covariance}
\State $c_{\Lambda}  = c_{\Lambda}  + q_1^{-1}$ \label{alg2:wedg1:end}
\EndFor
\vspace{0.5mm}
\State /*Compute wedge estimates for wedges $(v_3,v_2,v_1)$*/
\ForAll{$v_3 \in \hat \Gamma(v_2)$} \label{alg2:search_max_deg}
\State $k_2 \assign (v_2,v_3)$
\State $q_2 \assign \text{min}\{1,w(k_2)/z^*\}$
\State $\hat N_k(\Lambda) \pluseq (q q_2)^{-1}$ \Comment{wedge count} \label{alg2:wedg2:start}
\State $\hat V_k(\Lambda) \pluseq (q q_2)^{-1} ((q q_2)^{-1} -1)$ \Comment{wedge variance}
\State $\hat C_k(\Lambda) \pluseq c_{\Lambda} * q_2^{-1}$ \Comment{wedge covariance}
\State $c_{\Lambda}  = c_{\Lambda}  + q_2^{-1}$ \label{alg2:wedg2:end}
\EndFor
\State $\hat C_k(\vartriangle) = \hat C_k(\vartriangle) * 2 * q^{-1} * (q^{-1} - 1)$
\State $\hat C_k(\Lambda) = \hat C_k(\Lambda) * 2 * q^{-1} * (q^{-1} - 1)$
\EndParFor 
\State /*Compute total triangle/wedge estimates*/
\State $\hat N(\vartriangle) \assign \frac{1}{3} * \sum_{k \in \hat K} \hat N_k(\vartriangle)$, $\hat N(\Lambda)  \assign  \frac{1}{2} *  \sum_{k \in \hat K} \hat N_k(\Lambda)$ \label{alg2:reduce:start}
\State $\hat V(\vartriangle) \assign \frac{1}{3} * \sum_{k \in \hat K} \hat V_k(\vartriangle)$, $\hat V(\Lambda) \assign  \frac{1}{2} *  \sum_{k \in \hat K} \hat V_k(\Lambda)$
\State $\hat C(\vartriangle) \assign \sum_{k \in \hat K} \hat C_k(\vartriangle)$, $\hat C(\Lambda) \assign \sum_{k \in \hat K} \hat C_k(\Lambda)$
\State $\hat V(\vartriangle) \assign \hat V(\vartriangle) + \hat C(\vartriangle)$
\State $\hat V(\Lambda) \assign \hat V(\Lambda) + \hat C(\Lambda)$ \label{alg2:reduce:end}
\Return $\hat N(\vartriangle), \hat N(\Lambda), \hat V(\vartriangle), \hat V(\Lambda)$ \label{alg2:return}
\EndProcedure
\smallskip
\end{algorithmic}
\end{spacing}
\end{algorithm}
\end{minipage}
\vspace{-7mm}
\end{center}
\end{figure}
}
\section{In-Stream Estimation}
\label{sec-instream}
The foregoing analysis enables retrospective subgraph queries: 
after any number $t$ of stream
arrivals have taken place, we can compute an unbiased 
estimator $\hat S_t(J$) for any subgraph $J$. We term this \textbf{Post-Stream
  Estimation}. 
We now describe a second estimation framework that we call \textbf
{In-Stream Estimation}. In this paradigm, we
can take ``snapshots'' of specific sampled subgraphs at arbitrary times
during the stream, and preserve them as 
unbiased estimators. These can be used or combined to form
desired  estimators.
These snapshots are not subject to
further sampling; their estimates are not
updated. However their original subgraphs remain in the graph
sample and are subject to sampling in the normal way. Thus we do not
change the evolution of the graph sample, we only extract information
from it that does not change after extraction. 
The snapshot times need not be deterministic.
For example, each time a subgraph that matches a specified motif appears 
(e.g. a triangle or other clique) we take a snapshot of the subgraph 
estimator. If we only need to estimate the number of such subgraphs, 
it suffices to add the inverse probability of each matching subgraph to a counter. 

\begin{algorithm}[H]
\caption{
In-Stream Estimation of Triangle \& Wedge Counts
}
\label{alg:instream_gps_triangle}
\begin{spacing}{1.1}
\fontsize{8}{9}\selectfont
\algrenewcommand{\alglinenumber}[1]{\fontsize{6.5}{7}\selectfont#1}
\begin{algorithmic}[1]
\Procedure{\textsc{$\text{Instream\_GPS}$}}{$K$}
\State$\hat K \assign\emptyset$; $z^*\assign 0$ 
\While{new edge $k$}
\State \Call{GPSestimate}{$k$}
\State \Call{GPSupdate}{$k, m$}
\EndWhile
\Return $\tilde N(\vartriangle), \tilde N(\Lambda), \tilde V(\vartriangle),
\tilde V(\Lambda), \tilde V(\vartriangle,\Lambda)$
\EndProcedure
\vspace{-2mm}
\\\hrulefill
\Procedure{GPSestimate}{$k$}
\ParFor{Triangle $(k_1,k_2,k)$ completed by $k$}\label{alg3:par1}
\If{$(z^*==0)$}{\;$q_1\assign q_2 \assign 1$}
\Else
\State $q_1 \assign \min\{1,w(k_1)/z^*\}$
\State $q_2 \assign \min\{1,w(k_2)/z^*\}$
\EndIf
\State $\tilde N(\vartriangle)\pluseq 1/(q_1 q_2)$\Comment{Triangle Count}\label{alg3:nht}
\State $\tilde V(\vartriangle)\pluseq ( (q_1 q_2)^{-1}-1)/ (q_1 q_2)$\Comment{Triangle Var.}\label{alg3:sv}
\State $\tilde V(\vartriangle)\pluseq 2(\tilde C_{k_1}(\vartriangle) +
\tilde C_{k_2}(\vartriangle))/(q_1 q_2)$\label{alg3:cv}
\State $\tilde V(\vartriangle,\Lambda) \pluseq (\tilde C_{k_1}(\Lambda) +
\tilde C_{k_2}(\Lambda))/(q_1
q_2)$\Comment{Tri.-Wedge Cov.}\label{alg3:twc1}
\State $\tilde C_{k_1}(\vartriangle) \pluseq
(q_1^{-1}-1)/q_2$\Comment{Triangle Covariance}\label{alg3:cvup1}
\State $\tilde C_{k_2}(\vartriangle)  \pluseq (q_2^{-1}-1)/q_1$\label{alg3:cvup2}
\EndParFor

\ParFor{Wedge $j\in\hat K$ adjacent to $k$}\label{alg3:par2}
\If{$(z^*==0)$}{\;$q\assign  1$}
\Else{\;$q\assign \min\{1,w(j)/z^*\}$}
\EndIf
\State $\tilde N(\Lambda)\pluseq q^{-1}$\Comment{Wedge Count}\label{alg3:wcinc}
\State $\tilde V(\Lambda)\pluseq q^{-1} (q^{-1}
-1)$\Comment{Wedge Variance}\label{alg3:wvinc}
\State $\tilde V(\Lambda)\pluseq 2\tilde C_j(\Lambda)/q$
\State $\tilde V(\vartriangle,\Lambda) \pluseq \tilde C_j(\vartriangle)/q$\label{alg3:twc2}
\State $\tilde C_j(\Lambda)\pluseq 1/q -1$\Comment{Wedge Covariance}\label{alg3:par2:end}
\EndParFor
\EndProcedure
\vspace{-2mm}
\\\hrulefill
\Procedure{GPSupdate}{$k, m$}
  \State Generate $u(k)$ uniformly on $ (0,1]$
  \State $w(k)\assign W(k,\hat K)$
  \State $r(k)\assign w(k)/u(k)$ \Comment{Priority of edge $k$}
  \State $\hat K\assign \hat K\cup\{k\}$ \Comment{Provisionally include edge $k$}
  \State $\tilde C_k(\vartriangle)\assign \tilde C_k(\Lambda)\assign 0$
  \If{$|\hat K| > m$}
  \State $k^*\assign\argmin_{k'\in \hat K}r(k')$ \Comment{Lowest priority edge}
  \State $z^*\assign\max\{z^*,r(k^*)\}$ \Comment{New threshold}
  \State $\hat K \assign \hat K\setminus\{k^*\}$ \Comment{Remove lowest priority edge}
  \State $\tilde C(\vartriangle) \assign \tilde C(\vartriangle) \setminus
  \tilde C_{k^*}(\vartriangle)$ \Comment{Remove covariances of $k^*$}
 \State $\tilde C(\Lambda) \assign \tilde C(\Lambda) \setminus
  \tilde C_{k^*}(\Lambda)$

  \EndIf
\EndProcedure
\smallskip
\end{algorithmic}
\end{spacing}
\end{algorithm}  

\subsection{Unbiased Estimation with Snapshots}\label{sec:snap:unb}
In-stream estimation can be described within the framework of
\textsl{stopped Martingales}~\cite{W91}. Return for the moment
to our general description in Section~\ref{sec:framework} of a Martingale
$\{X_t: t\in \Nl\}$ with respect to a filtration $\cF=\{\cF_t: t\in\Nl\}$. A random time
$T$ is called a \textbf{stopping time} w.r.t $\cF$ if the event $T\le
t$ is $\cF_t$-measurable, i.e., we can decide at time $t$ whether the
event has occurred yet. The corresponding \textbf{stopped Martingale}
is
\be 
X^T= \{X^T_t: t\in\Nl\}\mbox{ where } X^T_t = X_{\min\{T,t\}}
\ee
Thus, the value of $X_t$ is frozen at $T$.

We define a \textbf{snapshot} as an edge subset $J\subset\Nl$
and a family $T=\{T_j:\ j \in J\}$ for $\cF_J$-measurable
stopping times, giving rise to the product stopped process 
\be\label{eq:snap:def}
\hat S^T_{J,t}=\prod_{j\in J}\hat S^{T_j}_{j,t}=  \prod_{j\in J}
\hat S_{j,\min\{T_j,t\}}
\ee
Although in-steam estimates use snapshots whose edges have the
same stopping time, variance estimation involves products of snapshots
with distinct stopping times. Unbiasedness then follows from the
following result that applies to any snapshot of the form (\ref{eq:snap:def}).

\begin{theorem}\label{thm:snap:unb} 
\begin{itemize}
\item[(i)] $\{\hat S^T_{J,t}:\ t\ge \max J\}$ is 
Martingale with respect to $\cF_J$ and hence $\E[\hat
S_{J,t}^{T}]=S_{J,t}$.
\item[(ii)] For any set $\cJ$ of subgraphs of $G$, each $J\subset \cJ$
  equipped with an $\cF_J$-measurable stoppong time $T_J$, then 
$\sum_{J\in \cT_j}\hat S^{T_J}_{J,t}$ is an unbiased estimator of
$|\cJ_t|$.
\end{itemize}
\end{theorem}

\subsection{Variance Estimation for Snapshots}\label{sec:snap:cov}

In this section we show how the variance and covariances of snapshots
can be estimated as sums of other snapshots involving the stopping
times of both constituents. The Martingale formulation is a powerful
tool to establish the unbiasedness of the estimates, since the
otherwise statistical properties of the product of correlated edge
estimators drawn at different times are not evident.

Consider two edge sets $J_1$ and $J_2$ each equipped
with families stopping times $T\up i = \{T\up i_j:\ j\in J_i\}$, with $i=1,2$,
for the purpose of snapshots. Thus each $j\in J_1\cap J_2$ is equipped
with two generally distinct stopping times $T\up 1_j$ and $T\up 2_j$,
according to its occurence in the snapshots $\hat S^{T\up 1}_{J_1,t}$ and
$\hat S^{T\up 2}_{J_2,t}$. As an estimator of $\cov( S^{T\up 1}_{J_1,t} ,
S^{T\up 2}_{J_2,t})$ we will use: 
\be\label{eq:cov:est:snap}
\hat C^{T\up 1,T\up 2}_{J_1,J_2,t}=\hat S^{T\up 1}_{J_1,t} \hat S^{T\up
  2}_{J_2,t} -\hat S^{T\up 1}_{J_1\setminus J_2,t} \hat S^{T\up
  2}_{J_2\setminus J_1,t}\hat S^{T\up 1\vee T\up 2}_{J_1\cap J_2,t}
\ee
where $T\up 1\vee T\up 2 = \{\max\{T\up 1_j,T\up 2_j\}:\ j\in J_1\cap
J_2\}$, i.e., we use the earlier stopping time for edges common to
both subsets.

\begin{theorem}\label{thm:snap:cov}
\begin{itemize}
\item[(i)] $\hat C^{T\up 1,T\up 2}_{J_1,J_2,t}$ is an unbiased
  estimator of  \break $\cov( \hat S^{T\up 1}_{J_1,t} , \hat S^{T\up
    2}_{J_2,t})$.
\item[(ii)] $\hat C^{T\up 1,T\up 2}_{J_1,J_2,t}\ge0$ and hence
  $\cov( \hat S^{T\up 1}_{J_1,t} , \hat S^{T\up 2}_{J_2,t})\ge0$.
\item[(iii)] $\hat S^{T}_{J,t}(\hat S^{T}_{J,t}-1)$ is an unbiased
  estimator of $\var(\hat S^{T}_{J,t})$.
\item[(iv)] $\hat C^{T\up 1,T\up 2}_{J_1,J_2,t}=0$ if and only if $\hat
  S^{T\up 1}_{J_1,t}=0$ or $\hat
  S^{T\up 2}_{J_2,t}=0$, i.e., covariance estimators are computed only
  from snapshots that have been sampled.
\end{itemize}
\end{theorem}

\parab{Covariance Estimation for Post-Stream Sampling.} Post-stream
sampling is a special case of in-stream sampling with all
$T_j=\infty$. We recover the corresponding Thm~\ref{thm:post:cov}
concerning covariances from
Theorem~\ref{thm:snap:cov} by omitting all stopping times $T_j$ from the notation.

\subsection{In-Stream Triangle \& Wedge Counts}

We now describe an application of In-Stream Estimation to Triangle
Sampling and Counting. We sample from the stream
based on the previous triangle count weighting,
but the triangle counting is
performed entirely in-stream. The full process is
described in Algorithm~\ref{alg:instream_gps_triangle}. In
this section we state and prove the form of estimators for triangle
count and its variance, and describe the corresponding steps in the
algorithm. Space limitations preclude giving a similar level of detail for
wedge counting and triangle-wedge covariance.

\parab{Unbiased In-Stream Estimation of Triangle Count.}
Using our previous notation $\vartriangle_t$ to denote the set of triangles all
of whose edges have arrived by $t$, we write each such triangle in the form
$(k_1,k_2,k_3)$ with $k_3$ the last edge to arrive.
If $k_1$ and $k_2$ are present in the
sample $\hat K$ when $k_3$ arrives, we take a snapshot of the wedge 
$(k_1,k_2)$ \textit{prior to the sampling step for $k_3$}. Formally,
we let $T_{k_3}$ denote the slot immediately prior the arrival of
$k_3$ and form the snapshot $\hat
S^{T_{k_3}}_{\{k_1,k_2\}}$. Note $T_{k_3}$ is a stopping time because
the edge arrival process is deterministic.

\begin{theorem}\label{thm:tri:instream}
$\tilde N_t(\vartriangle) = \sum_{(k_1,k_2,k_3)\in\vartriangle_t} \hat S^{T_{k_3}}_{\{k_1,k_2\},t}$ is an
unbiased estimator of $N_t(\vartriangle )$.
\end{theorem}

\parab{Estimation of Triangle Count Variance.}
We add some further notation in preparation for estimating the
variance of $\tilde N_t(\vartriangle)$. Let $\hat K_{[t]} = \cup_{t>0}\hat K_t$ denote the set of
  edges that are sampled at any time up to $t$. Let $\hat \vartriangle_t=\{(k_1,k_2,k_3)\in \vartriangle_t: \hat
  S_{\{k_1,k_2\}}^{T_{k_3}}>0$\} denote the (random) subset of all triangles in
  $\vartriangle_t$ that have positive
  snapshot.  Let $\hat K\up 2_{[t]}(k)$ denote the
    set of pairs $(j',k')$ of edges in $\hat K_{[t]}$ such that each of
    the edge pairs $(k,j')$
    and $(k,k')$ are the first two edges in distinct triangles in
    $\hat\vartriangle_t$, and with $(j',k')$ ordered by their
    stopping time of the third edges in these triangles, i.e.,  $T_{k,j'}<T_{k,k'}$.

\begin{theorem}\label{thm:tri:instream:var} 
$\var(\tilde N_t(\vartriangle))$ has unbiased estimator
\bea
\tilde V_t(\vartriangle)=\kern -10pt \sum_{(k_1,k_2,k_3)\in \hat \vartriangle_t} \frac{1}{p_{k_1,T_{k_3}}
  p_{k_2,T_{k_3}}} \left(  \frac{1}{p_{k_1,T_{k_3}}
  p_{k_2,T_{k_3}}} -1\right)\kern -220pt && \\
&& \kern -10pt +2 \sum_{ k\in\hat K_{[t]}}
\sum_{{(j',k')\in \atop \hat K\up 2_{[t]}(k)}}
\frac{1}{p_{k',T_{k,k'}} p_{j',T_{k,j'}}p_{k,T_{k,k'}}}
\left(
\frac{1}{p_{k,T_{k,j'}}} -1
\right)\nonumber
\eea  
\end{theorem}

\parab{Description of Algorithm~\ref{alg:instream_gps_triangle}.}
We now describe the portions of
Algorithm~\ref{alg:instream_gps_triangle} relating the in-stream
estimator $\tilde N(\vartriangle)$ and $\tilde V(\vartriangle)$. When
an edge $k$ arrives, an update is performed for each triangle 
that $k$ completes (line ~\ref{alg3:par1}). These updates can be performed in
parallel because each such triangle must have distinct edges other
than $k$. Triangle count $\tilde N(\vartriangle)$ is updated with the
current inverse probabilities of its first two edges $k_1$ and $k_2$
(line~\ref{alg3:nht}). The variance $\tilde V(\vartriangle)$ is updated
first with the variance term for the current triangle (line \ref{alg3:sv}) then secondly with
its cumulative contributions $\hat C_{k_1}(\vartriangle)$ and $\hat
C_{k_1}(\vartriangle)$ to the covariances with all previous triangles whose first two
edges include $k_1$ or $k_2$ (line~\ref{alg3:cv}). These cumulative
terms are then updated by the current triangle (lines ~\ref{alg3:cvup1}
and ~\ref{alg3:cvup2}). Wedge count variables are updated in a similar
manner in lines \ref{alg3:par2}--\ref{alg3:par2:end}. The
edge-wedge covariance $\tilde V(\vartriangle,\Lambda)$ used for
estimation of the global clustering coefficient $\alpha$ is updated
using the cumulative triangle and wedge covariances in lines \ref{alg3:twc1}
and \ref{alg3:twc2}.

\section{Experiments and Results}
\label{sec:eval}

We test the performance of graph priority sampling on a large set of $50$ graphs with hundreds of millions of nodes/edges, selected from a variety of domains and types, such as social, web, among others. All graph data sets are available for download~\footnote{$\mathtt{www.networkrepository.com}$}~\cite{nr-aaai15,nr-sigkdd16}.
For all graph datasets, we consider an undirected, unweighted, simplified graph without self loops. We generate the graph stream by randomly permuting the set of edges in each graph. For each graph, we perform ten different experiments with sample sizes in the range of 10K--1M edges. We test \gps\ as well as baseline methods using a single pass over the edge stream (such that each edge is processed once): both \gps\ post and in-stream estimation randomly select the same set of edges with the same random seeds. Thus, the two methods only differ in the estimation procedure. For these experiments, we used a server with two Intel Xeon E5-2687W 3.1GHz CPUs. Each processor has 8 cores with 20MB of L3 cache and 256KB of L2 cache. The experimental setup is executed independently for each sample size as follows:

\begin{enumerate}[leftmargin=0pt,itemindent=10pt,itemsep=1pt,parsep=0pt,topsep=1pt,partopsep=0pt]
\item For each sample size $m$, Alg~\ref{alg:alg1} collects an edge sample $\hat K \subset K$.
\item We use Alg~\ref{alg:alg_gpsest} for post stream estimation, where the estimates are obtained only from the sample. We use Alg~\ref{alg:instream_gps_triangle} for in-stream estimation, where the estimates are updated incrementally in a single pass during the sampling process. Thus, both \gps\ post and in-stream estimation use the same sample.
\item Given a sample $\hat K \subset K$, we use the absolute relative error (ARE) $|\E[\hat X] - X|/X$ to measure the deviation between the expected value of the estimates $\hat X$ and the actual statistics $X$. We use $X$ to refer to triangle, wedge counts, or global clustering.
\item We compute the $95\%$ confidence bounds as $\hat X \pm 1.96 \sqrt{\var[\hat X]}$.
\end{enumerate}

\begin{table*}[t!]
\scriptsize
\caption{Estimates of expected value and relative error using 200K edges for a representative set of 11 graphs.
The graphlet statistic for the full graph is shown in the third column.
LB and UB are 95\% confidence lower and upper bounds, respectively.
}
\tiny
\vspace{1mm}
\label{table:est-200k}
\centering\small\scriptsize
\begin{tabularx}{1.0\linewidth}{@{}p{5mm}@{}r@{} c HlX X XXHXX c XXHXX @{}}

\toprule
&&&
&&&
{\sc \fontsize{8.5}{9.5}\selectfont actual}  &
\multicolumn{5}{c}{{\sc \fontsize{8.5}{9.5}\selectfont \gps\ in-stream}}&&
\multicolumn{5}{c}{{\sc \fontsize{8.5}{9.5}\selectfont \gps\ post stream}}
\\
\cmidrule(l{-3pt}r{5pt}){7-7}
\cmidrule(l{-0pt}r{3pt}){8-12}
\cmidrule(l{-3pt}r{0pt}){14-18}
& \textbf{graph}  &&  
$|V|$ & $|K|$ &  $\frac{|\hat K|}{|K|}$  & 
$X$  &  
$\widehat{X}$  &  $\frac{|X - \widehat{X}|}{X}$   &   $|\hat K|$   &    LB    &     UB    &&  
$\widehat{X}$  &  $\frac{|X - \widehat{X}|}{X}$   &   $|\hat K|$   &    LB    &     UB
\\
\midrule
\multirow{11}{*}{\rotatebox[origin=c]{90}{\mbox{
\rotatebox[origin=l]{315}{\includegraphics[scale=0.05]{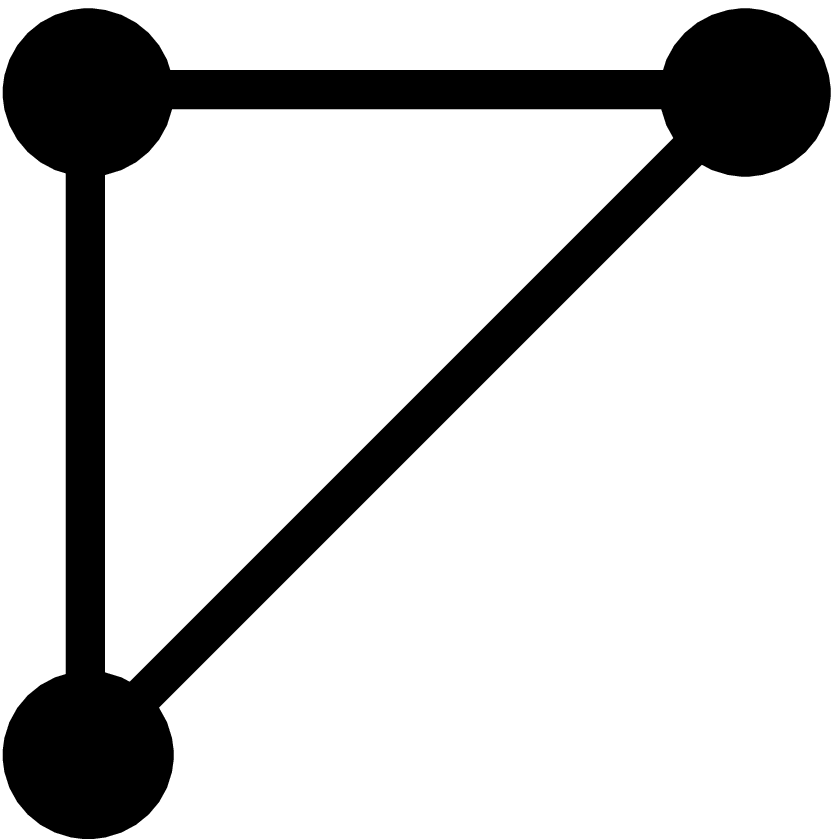} \;}
{\sc triangles }}}}
& \dataName{ca-hollywood-2009} && 
1M  &  56.3M  &  0.0036  &
4.9B   &
4.9B    &    \textbf{0.0009}    &   200000   &    4.8B    &    5B &&
4.8B    &    \textbf{0.0036}    &   200000   &    4.6B    &    5.1B 
\\

& \dataName{com-amazon} && 
334.8K  &  925.8K  &  0.216  &
667.1K   &
667.2K    &    \textbf{0.0001}    &   200000   &    658.5K    &    675.8K &&
666.8K    &    \textbf{0.0004}    &   200000   &    653.6K    &    680K 
\\

& \dataName{higgs-social-network} && 
456.6K  &  12.5M  &  0.016  &
83M   &
82.6M    &    \textbf{0.0043}    &   200000   &    80.8M    &    84.4M &&
83.2M    &    \textbf{0.0031}    &   200000   &    79.5M    &    87M 
\\

& \dataName{soc-livejournal} && 
4M  &  27.9M  &  0.0072  &
83.5M   &
83.1M    &    \textbf{0.0043}    &   200000   &    80.6M    &    85.7M &&
81.5M    &    \textbf{0.0244}    &   200000   &    72M    &    91M 
\\

& \dataName{soc-orkut} && 
3M  &  117.1M  &  0.0017  &
627.5M   &
625.8M    &    \textbf{0.0028}    &   200000   &    601.4M    &    650.1M &&
614.8M    &    \textbf{0.0203}    &   200000   &    396M    &    833.7M 
\\

& \dataName{soc-twitter-2010} && 
21.2M  &  265M  &  0.0008  &
17.2B   &
17.3B    &    \textbf{0.0009}    &   200000   &    16.8B    &    17.7B &&
17.3B    &    \textbf{0.0027}    &   200000   &    13.3B    &    21.3B 
\\

& \dataName{soc-youtube-snap} && 
1.1M  &  2.9M  &  0.0669  &
3M   &
3M    &    \textbf{0.0004}    &   200000   &    2.9M    &    3.1M &&
3M    &    \textbf{0.0003}    &   200000   &    2.9M    &    3.1M 
\\

& \dataName{socfb-Penn94} && 
41.5K  &  1.3M  &  0.1468  &
7.2M   &
7.1M    &    \textbf{0.0063}    &   200000   &    7M    &    7.2M &&
7.1M    &    \textbf{0.0044}    &   200000   &    6.8M    &    7.5M 
\\

& \dataName{socfb-Texas84} && 
36.3K  &  1.5M  &  0.1257  &
11.1M   &
11.1M    &    \textbf{0.0011}    &   200000   &    10.9M    &    11.3M &&
11.1M    &    \textbf{0.0013}    &   200000   &    10.4M    &    11.9M 
\\

& \dataName{tech-as-skitter} && 
1.6M  &  11M  &  0.018  &
28.7M   &
28.5M    &    \textbf{0.0081}    &   200000   &    27.7M    &    29.3M &&
28.3M    &    \textbf{0.0141}    &   200000   &    26.5M    &    30.1M 
\\

& \dataName{web-google} && 
875.7K  &  4.3M  &  0.0463  &
13.3M   &
13.4M    &    \textbf{0.0034}    &   200000   &    13.2M    &    13.6M &&
13.4M    &    \textbf{0.0078}    &   200000   &    13.1M    &    13.8M 
\\

\midrule
\multirow{11}{*}{\rotatebox[origin=c]{90}{\mbox{
\rotatebox[origin=l]{315}{\includegraphics[scale=0.05]{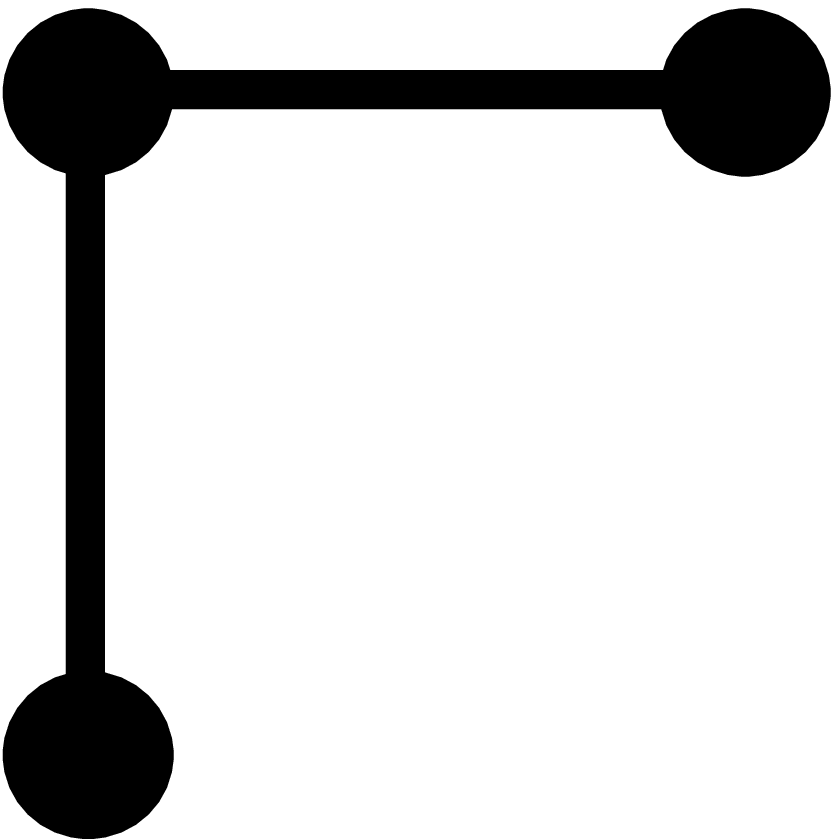} \;}
{\sc wedges }}}}
& \dataName{ca-hollywood-2009} && 
1M  &  56.3M  &  0.0036  &
47.6B   &
47.5B    &    \textbf{0.0011}    &   200000   &    47.3B    &    47.8B &&
47.5B    &    \textbf{0.0026}    &   200000   &    46.9B    &    48.1B 
\\

& \dataName{com-amazon} && 
334.8K  &  925.8K  &  0.216  &
9.7M   &
9.7M    &    \textbf{0.0002}    &   200000   &    9.7M    &    9.8M &&
9.7M    &    \textbf{0.0021}    &   200000   &    9.6M    &    9.9M 
\\

& \dataName{higgs-social-network} && 
456.6K  &  12.5M  &  0.016  &
28.7B   &
28.7B    &    \textbf{0.001}    &   200000   &    28.5B    &    28.9B &&
28.7B    &    \textbf{0.0008}    &   200000   &    28.1B    &    29.3B 
\\

& \dataName{soc-livejournal} && 
4M  &  27.9M  &  0.0072  &
1.7B   &
1.7B    &    \textbf{0.0005}    &   200000   &    1.7B    &    1.8B &&
1.8B    &    \textbf{0.0008}    &   200000   &    1.7B    &    1.8B 
\\

& \dataName{soc-orkut} && 
3M  &  117.1M  &  0.0017  &
45.6B   &
45.5B    &    \textbf{0.0016}    &   200000   &    45B    &    46B &&
45.5B    &    \textbf{0.0009}    &   200000   &    44.3B    &    46.8B 
\\

& \dataName{soc-twitter-2010} && 
21.2M  &  265M  &  0.0008  &
1.8T   &
1.8T    &    \textbf{0.0002}    &   200000   &    1.8T    &    1.8T &&
1.8T    &    \textbf{0.0016}    &   200000   &    1.7T    &    1.8T 
\\

& \dataName{soc-youtube-snap} && 
1.1M  &  2.9M  &  0.0669  &
1.4B   &
1.4B    &    \textbf{0.0035}    &   200000   &    1.4B    &    1.4B &&
1.4B    &    \textbf{0.0084}    &   200000   &    1.4B    &    1.5B 
\\

& \dataName{socfb-Penn94} && 
41.5K  &  1.3M  &  0.1468  &
220.1M   &
219.9M    &    \textbf{0.001}    &   200000   &    217.7M    &    222.1M &&
219M    &    \textbf{0.0051}    &   200000   &    211.7M    &    226.3M 
\\

& \dataName{socfb-Texas84} && 
36.3K  &  1.5M  &  0.1257  &
335.7M   &
334.9M    &    \textbf{0.0022}    &   200000   &    331.4M    &    338.5M &&
335.1M    &    \textbf{0.0017}    &   200000   &    323M    &    347.2M 
\\

& \dataName{tech-as-skitter} && 
1.6M  &  11M  &  0.018  &
16B   &
16B    &    \textbf{0.0005}    &   200000   &    15.8B    &    16.1B &&
15.9B    &    \textbf{0.0016}    &   200000   &    15.6B    &    16.3B 
\\

& \dataName{web-google} && 
875.7K  &  4.3M  &  0.0463  &
727.4M   &
728.8M    &    \textbf{0.002}    &   200000   &    721M    &    736.7M &&
732.2M    &    \textbf{0.0066}    &   200000   &    711.8M    &    752.5M 
\\

\midrule
\multirow{7}{*}{\rotatebox[origin=c]{90}{\mbox{
{\sc clustering coeff. (cc) }}}}
& \dataName{ca-hollywood-2009} && 
1M  &  56.3M  &  0.0036  &
0.31   &
0.31    &    \textbf{0.002}    &   200000   &    0.306    &    0.315 &&
0.309    &    \textbf{0.0009}    &   200000   &    0.295    &    0.323 
\\

& \dataName{com-amazon} && 
334.8K  &  925.8K  &  0.216  &
0.205   &
0.205    &    $\mathbf{<}$\textbf{10}$^{\mathbf{-4}}$    &   200000   &    0.203    &    0.208 &&
0.205    &    \textbf{0.0025}    &   200000   &    0.201    &    0.209 
\\

& \dataName{higgs-social-network} && 
456.6K  &  12.5M  &  0.016  &
0.009   &
0.009    &    \textbf{0.0034}    &   200000   &    0.008    &    0.009 &&
0.009    &    \textbf{0.0039}    &   200000   &    0.008    &    0.009 
\\

& \dataName{soc-livejournal} && 
4M  &  27.9M  &  0.0072  &
0.139   &
0.139    &    \textbf{0.0039}    &   200000   &    0.135    &    0.143 &&
0.136    &    \textbf{0.0252}    &   200000   &    0.12    &    0.151 
\\

& \dataName{soc-orkut} && 
3M  &  117.1M  &  0.0017  &
0.041   &
0.041    &    \textbf{0.0012}    &   200000   &    0.04    &    0.043 &&
0.04    &    \textbf{0.0193}    &   200000   &    0.026    &    0.055 
\\

& \dataName{soc-twitter-2010} && 
21.2M  &  265M  &  0.0008  &
0.028   &
0.028    &    \textbf{0.0012}    &   200000   &    0.028    &    0.029 &&
0.028    &    \textbf{0.0004}    &   200000   &    0.022    &    0.035 
\\

& \dataName{soc-youtube-snap} && 
1.1M  &  2.9M  &  0.0669  &
0.006   &
0.006    &    \textbf{0.0032}    &   200000   &    0.006    &    0.006 &&
0.006    &    \textbf{0.0088}    &   200000   &    0.006    &    0.007 
\\

& \dataName{socfb-Penn94} && 
41.5K  &  1.3M  &  0.1468  &
0.098   &
0.098    &    \textbf{0.0053}    &   200000   &    0.096    &    0.099 &&
0.098    &    \textbf{0.0008}    &   200000   &    0.093    &    0.104 
\\

& \dataName{socfb-Texas84} && 
36.3K  &  1.5M  &  0.1257  &
0.1   &
0.1    &    \textbf{0.0012}    &   200000   &    0.098    &    0.102 &&
0.1    &    \textbf{0.0031}    &   200000   &    0.093    &    0.107 
\\

& \dataName{tech-as-skitter} && 
1.6M  &  11M  &  0.018  &
0.005   &
0.005    &    \textbf{0.0076}    &   200000   &    0.005    &    0.006 &&
0.005    &    \textbf{0.0124}    &   200000   &    0.005    &    0.006 
\\

& \dataName{web-google} && 
875.7K  &  4.3M  &  0.0463  &
0.055   &
0.055    &    \textbf{0.0014}    &   200000   &    0.054    &    0.056 &&
0.055    &    \textbf{0.0013}    &   200000   &    0.053    &    0.057 
\\
\bottomrule
\end{tabularx}
\end{table*}

\parab{Error Analysis and Confidence Bounds.} Table~\ref{table:est-200k} summarizes the main graph properties and provides a comparison between \gps\ post stream and in-stream estimation for a variety of graphs at sample size $m =$ 200K edges. First, we observe that \gps\ in-stream estimation has on average $< 1\%$ relative error across most graphs. In addition, \gps\ post stream estimation has on average $\leq 2\%$. Thus, both methods provide accurate estimates for large graphs with a small sample size. Table~\ref{table:est-200k} also shows that the upper and lower bounds of \gps\ in-stream estimation are smaller than those obtained using \gps\ post stream estimation. We note that both methods are using the same sample. However, a key advantage for \gps\ in-stream estimation versus \gps\ post stream estimation is its ability to minimize the variance of the estimators. Thus, \gps\ in-stream estimates are not only accurate but also have a small variance and tight confidence bounds.

\begin{figure}
\vspace{-2mm}
\centering
\includegraphics[width=0.85\linewidth]{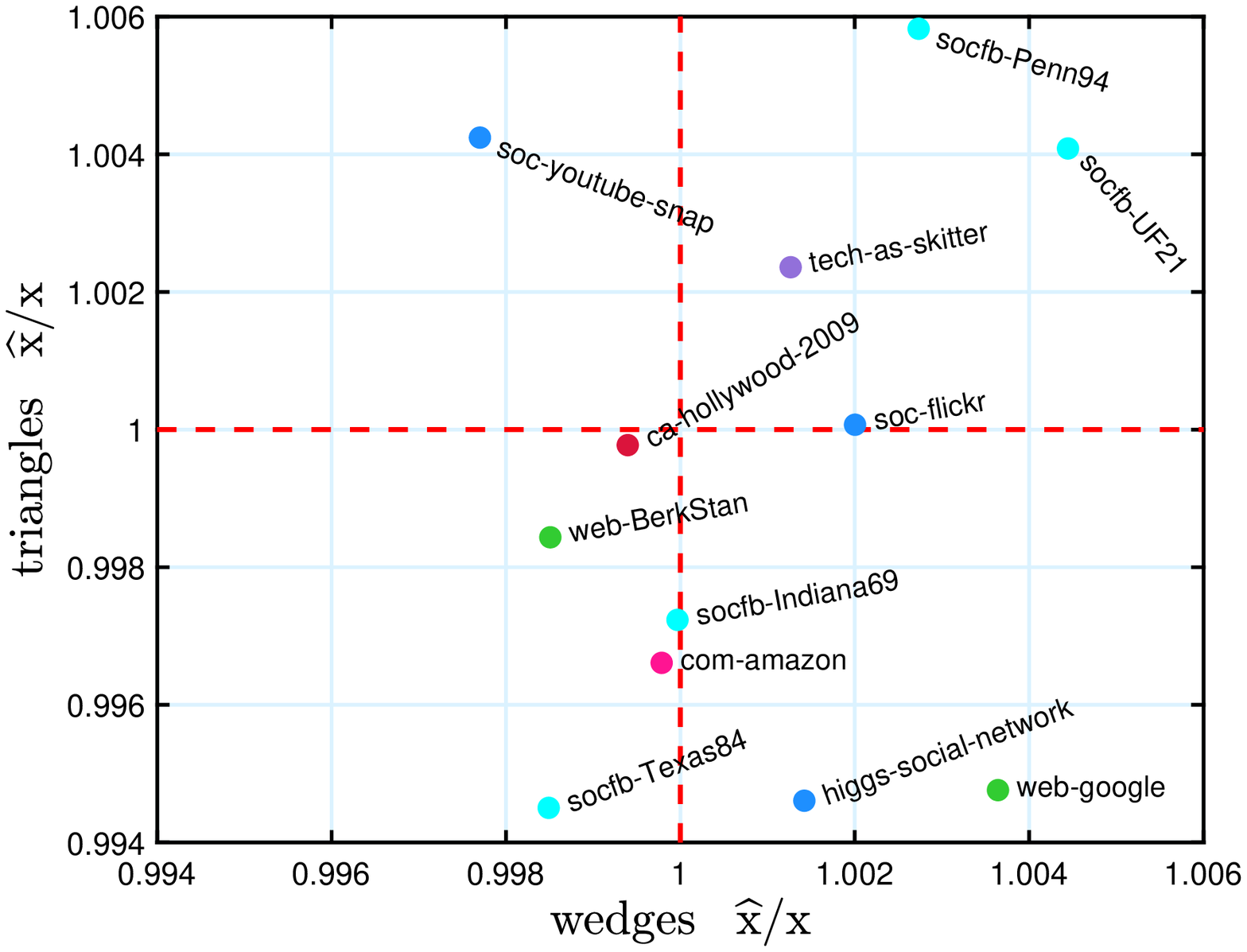}
\vspace{-3mm}
\caption{
Comparing $\nicefrac{\widehat{x}}{x}$ of triangles and wedges.
The closer the points are to the intersection of the red lines (actual) the better.
Points are colored by graph type. 
Results are from the in-stream estimation method at 100K.}
\label{fig:tri-vs-wedges-100k-instream}
\vspace{-2mm}
\end{figure}

Second, we observe that the \gps\ framework provides high quality general samples to accurately estimate various properties simultaneously. For example, Table~\ref{table:est-200k} shows consistent performance across all graphs for the estimation of triangle/wedge counts and global clustering with the same sample. Similarly, in Figure~\ref{fig:tri-vs-wedges-100k-instream}, we observe that \gps\ accurately estimates both triangle and wedge counts simultaneously with a single sample, with a relative error of $0.6\%$ for for both triangle and wedge counts. 

Finally, we investigate the properties of the sampling distribution and the convergence of the estimates as the sample size increases between 10K--1M edges (See Figure~\ref{fig:gps-triangle-confidence-bounds}). We used graphs from various domains and types. We observe that the confidence intervals of triangle counts are small in the range $0.90$--$1.10$ for most graphs at 40K sample size. Notably, for a large Twitter graph with more than 260M edges (soc-twitter-10), \gps\ in-stream estimation accurately estimates the triangle count with $<1\%$ error, while storing 40K edges, which is only a fraction of $0.00015$ of the total number of edges in the graph. Due to space limitations, we removed the confidence plots for wedges and clustering coefficient. However, we observe that the confidence interval are very small in the range of $0.98$--$1.02$ for wedges, and $0.90$--$1.10$ for global clustering coefficient. 


\providecommand{\figConfhspace}{-1mm}
\providecommand{\figConfSZ}{0.25\linewidth}

\begin{figure*}[t!]
\vspace{-4mm}
\centering

\hspace*{\figConfhspace}\subfigure	
{\includegraphics[width=\figConfSZ]{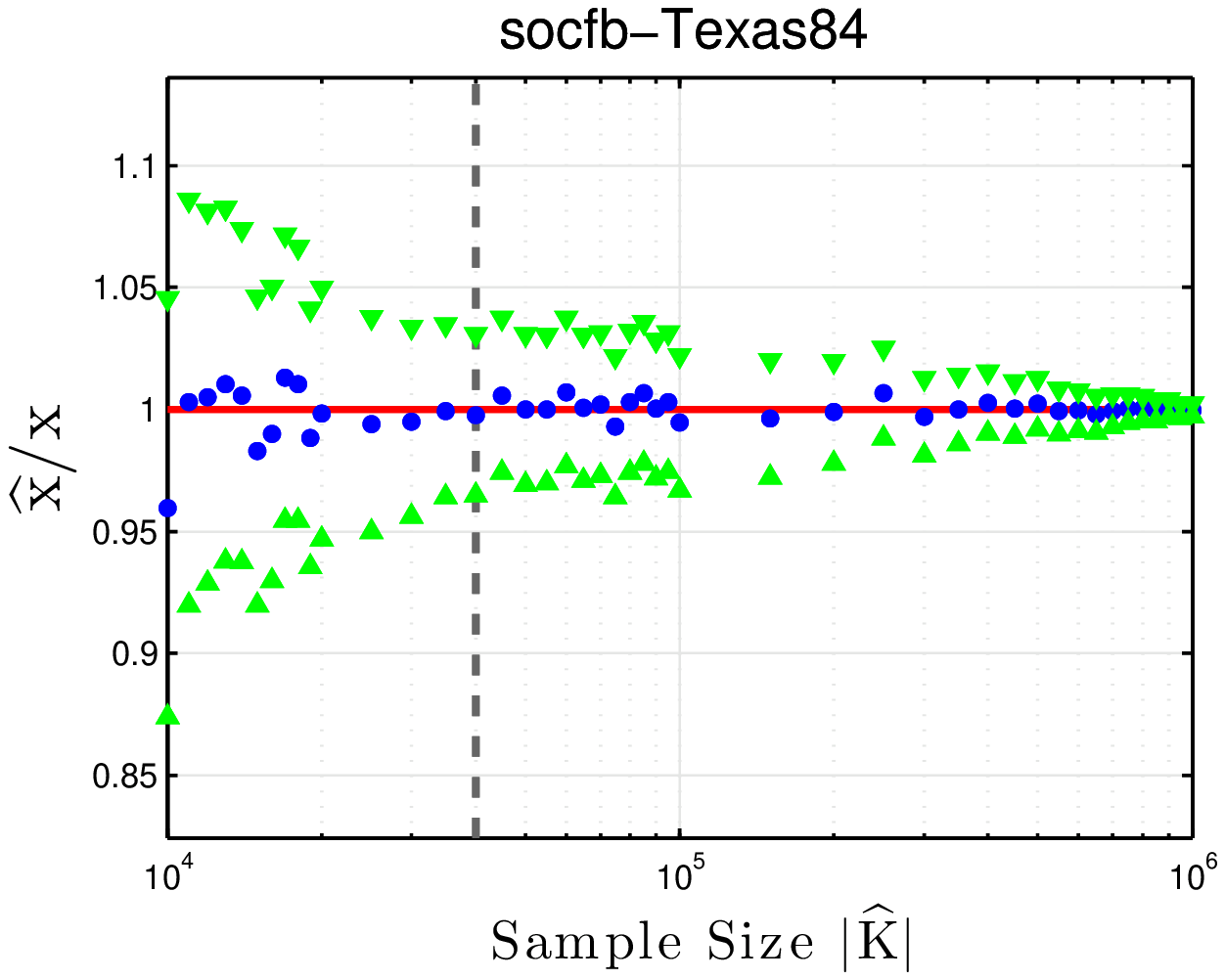}}
\hspace*{\figConfhspace}\subfigure	
{\includegraphics[width=\figConfSZ]{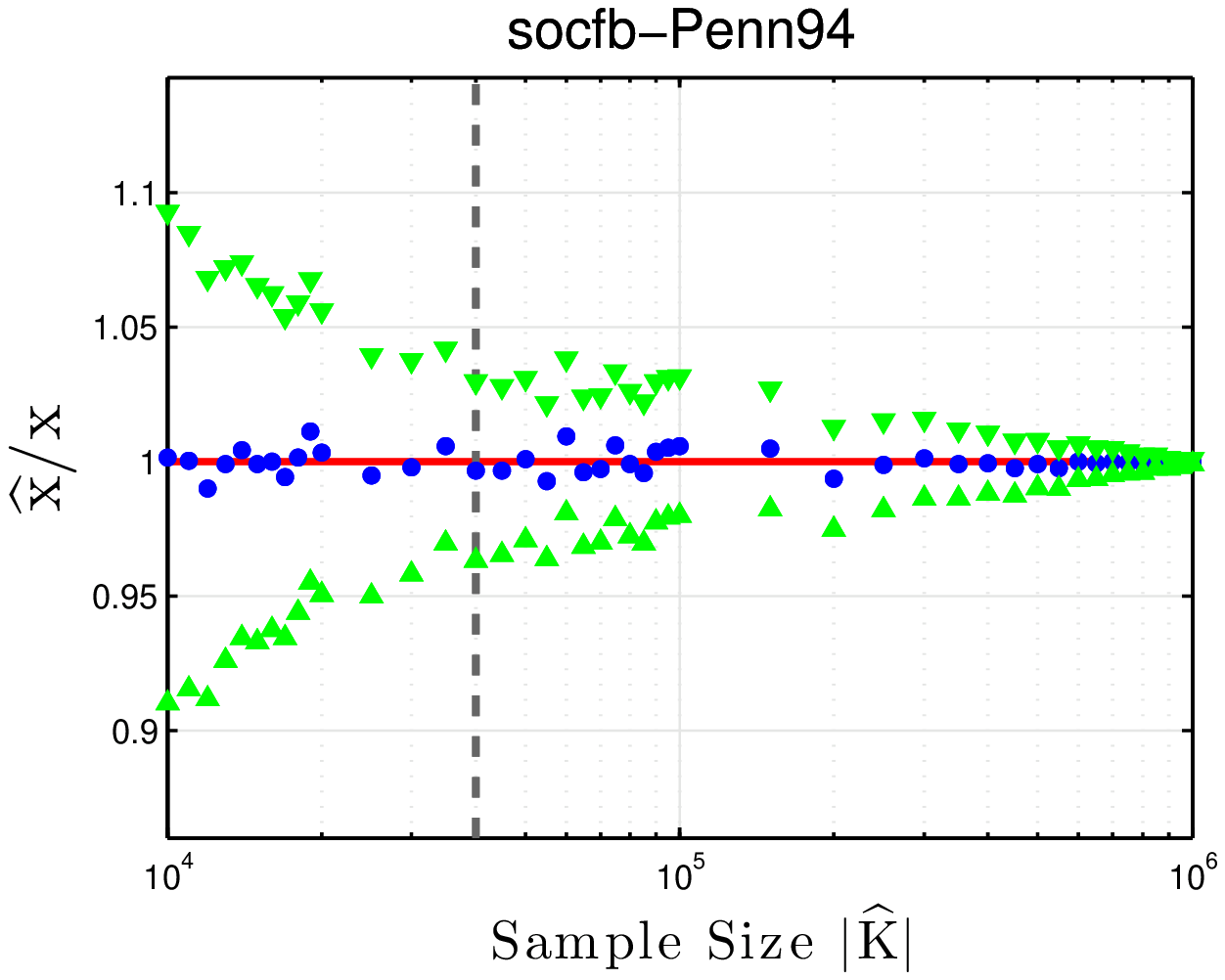}}
\hspace*{\figConfhspace}\subfigure	
{\includegraphics[width=\figConfSZ]{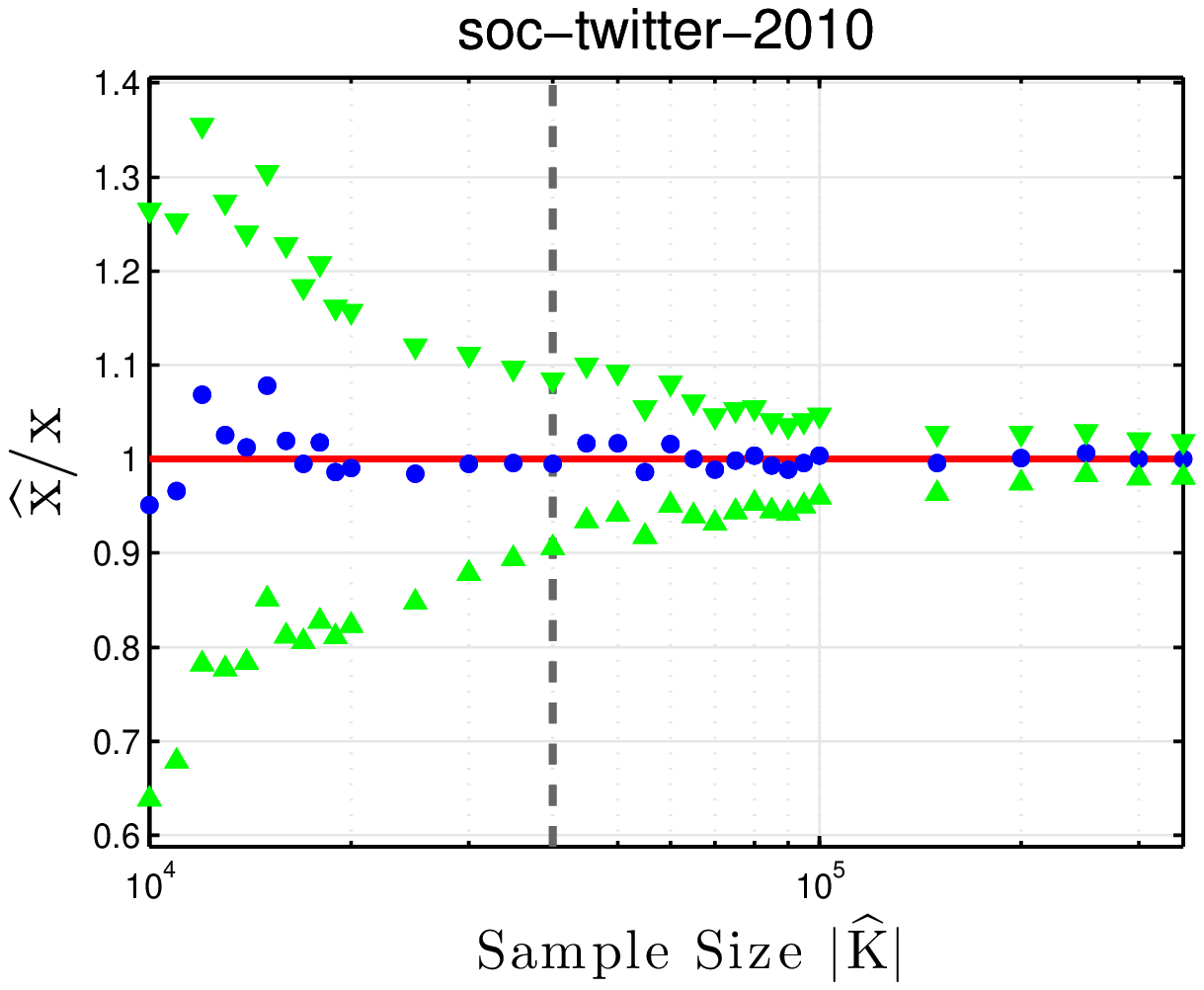}}
\hspace*{\figConfhspace}\subfigure		
{\includegraphics[width=\figConfSZ]{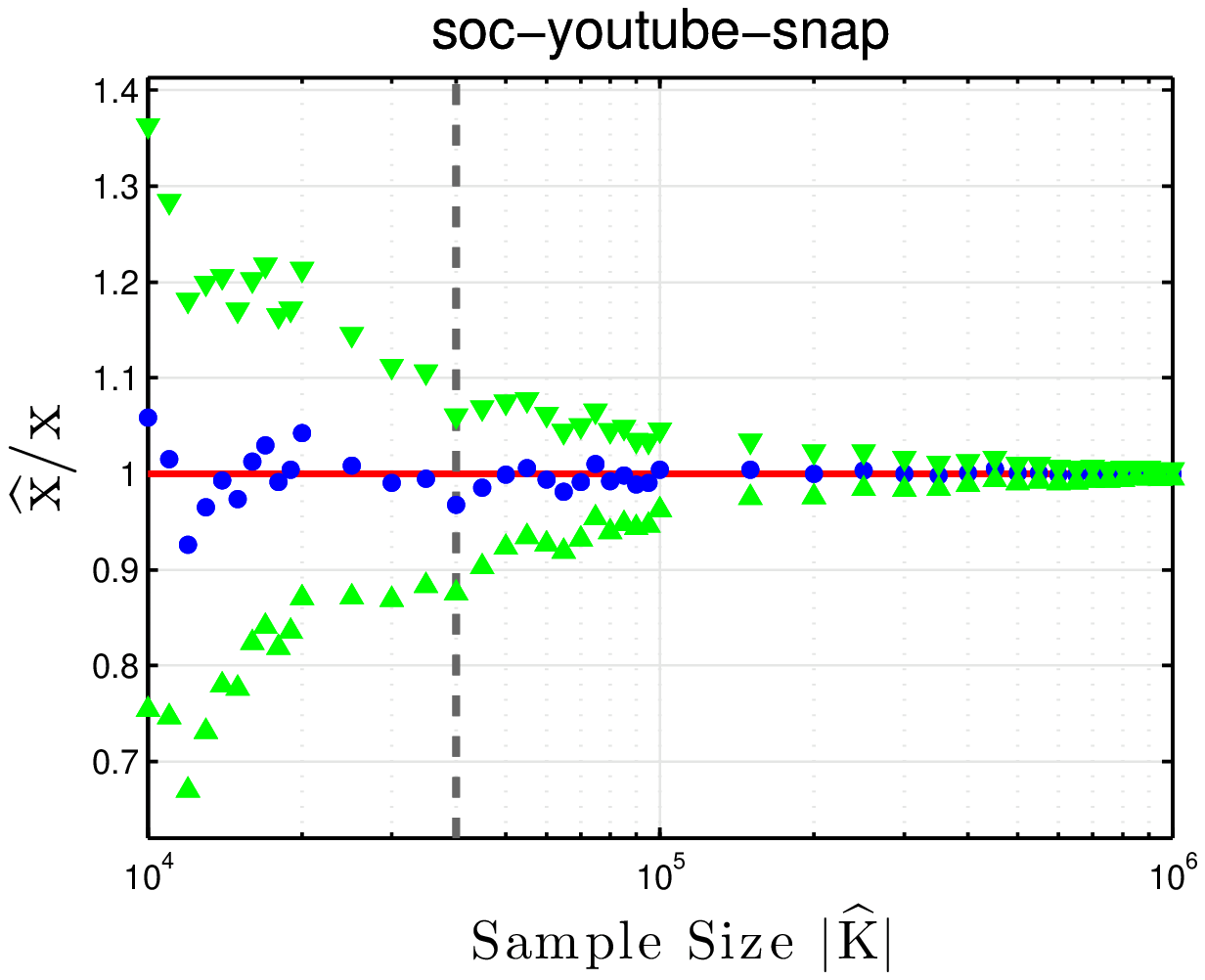}}

\hspace*{\figConfhspace}\subfigure	
{\includegraphics[width=\figConfSZ]{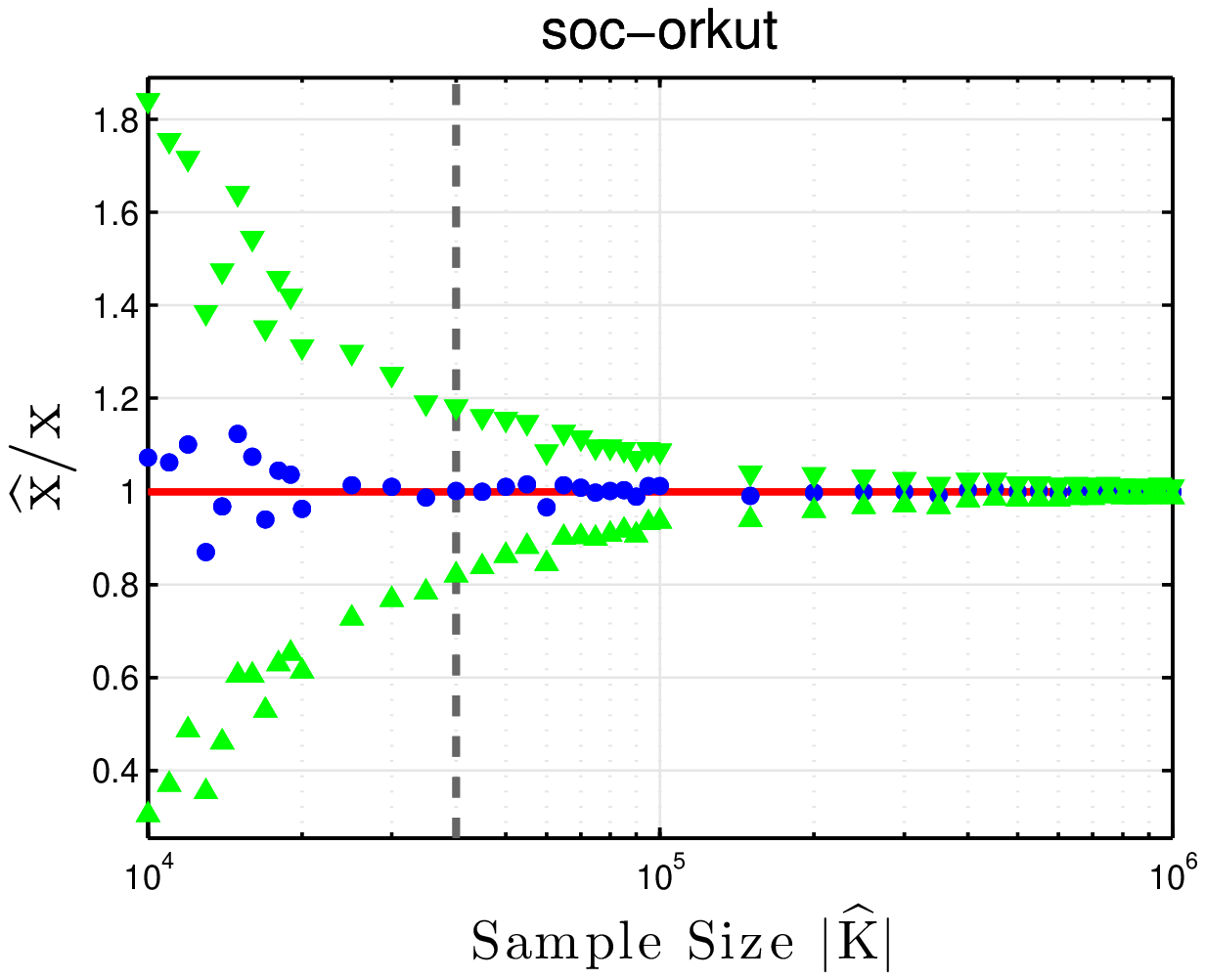}}
\hspace*{\figConfhspace}\subfigure		
{\includegraphics[width=\figConfSZ]{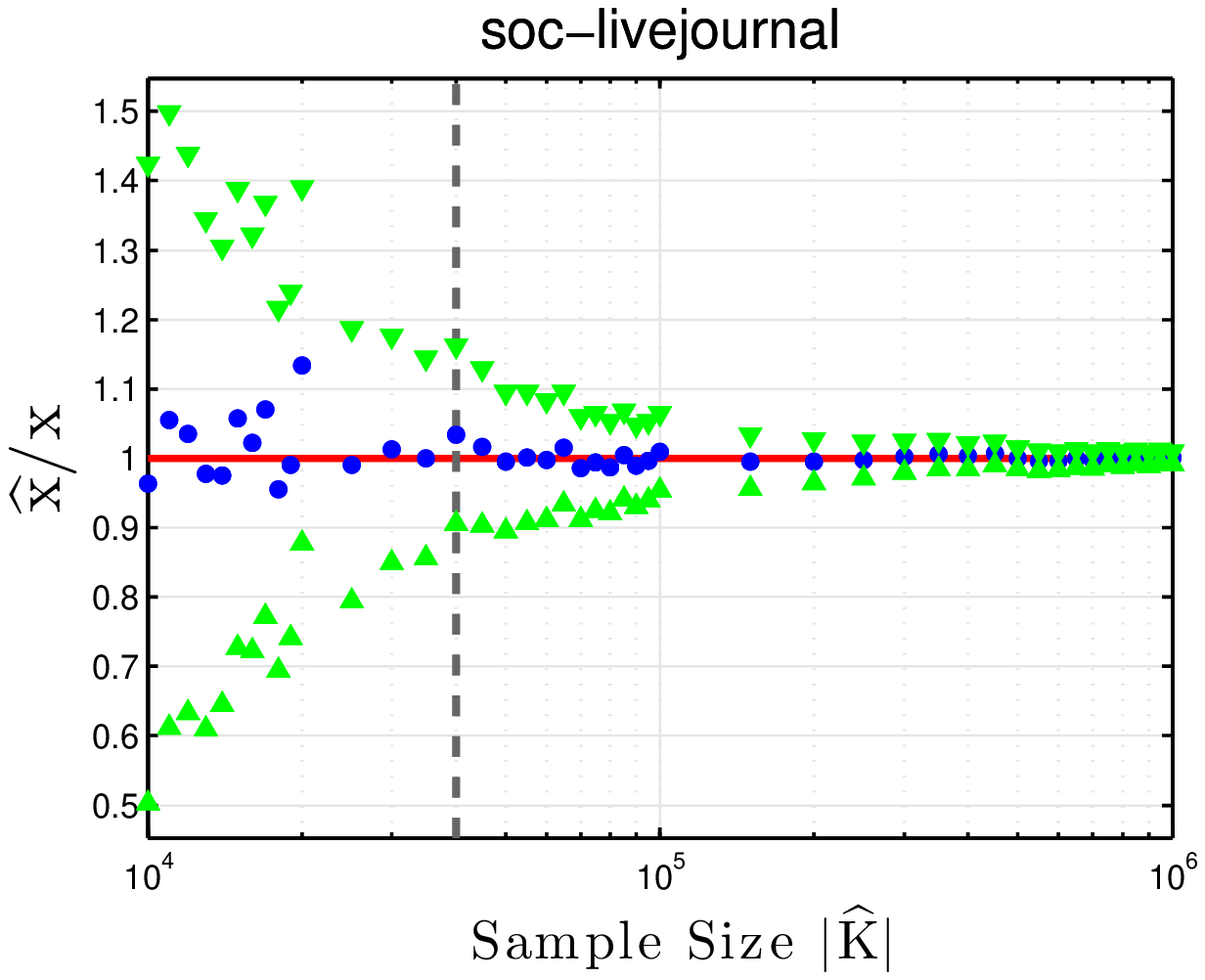}}
\hspace*{\figConfhspace}\subfigure		
{\includegraphics[width=\figConfSZ]{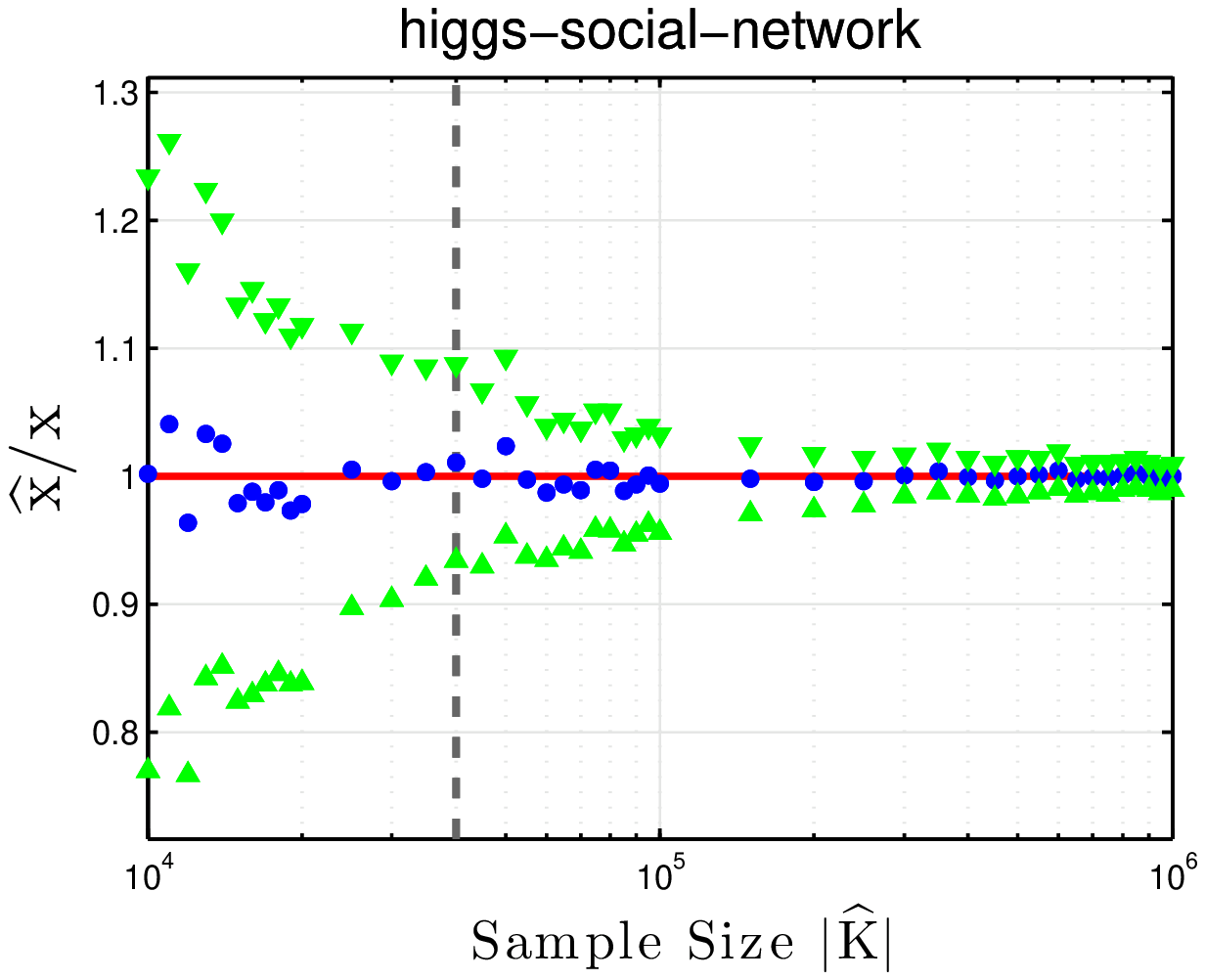}}
\hspace*{\figConfhspace}\subfigure	
{\includegraphics[width=\figConfSZ]{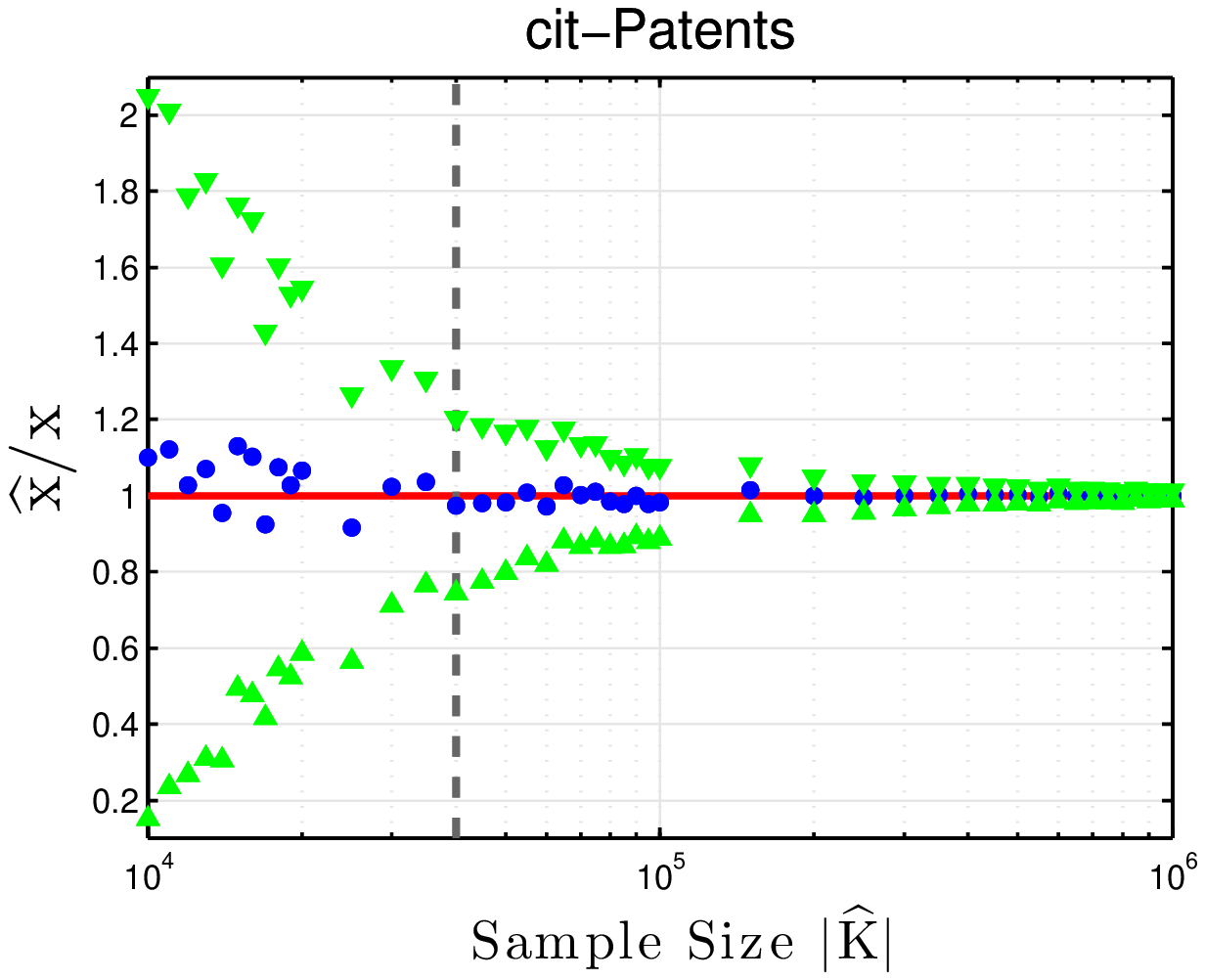}}

\hspace*{\figConfhspace}\subfigure		
{\includegraphics[width=\figConfSZ]{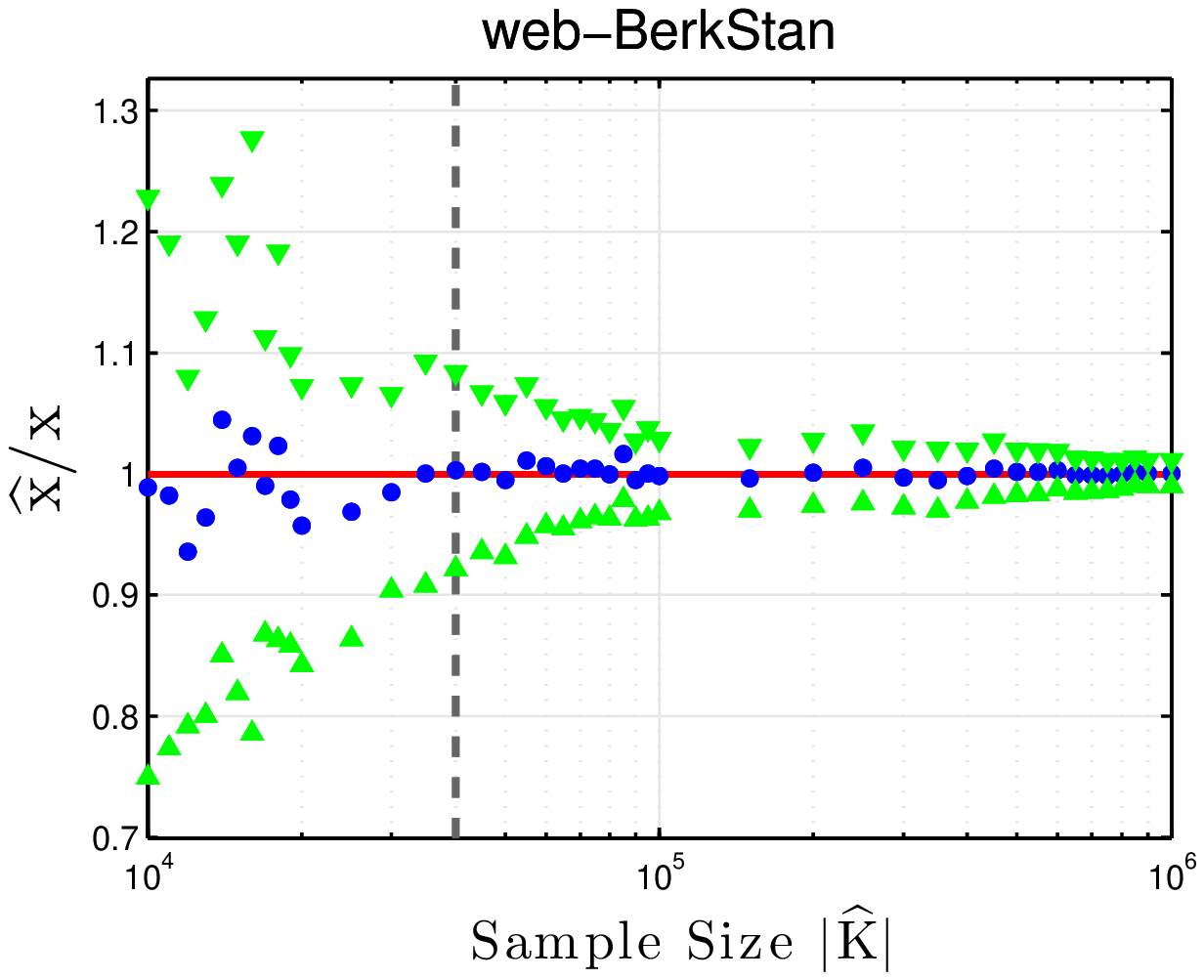}}
\hspace*{\figConfhspace}\subfigure		
{\includegraphics[width=\figConfSZ]{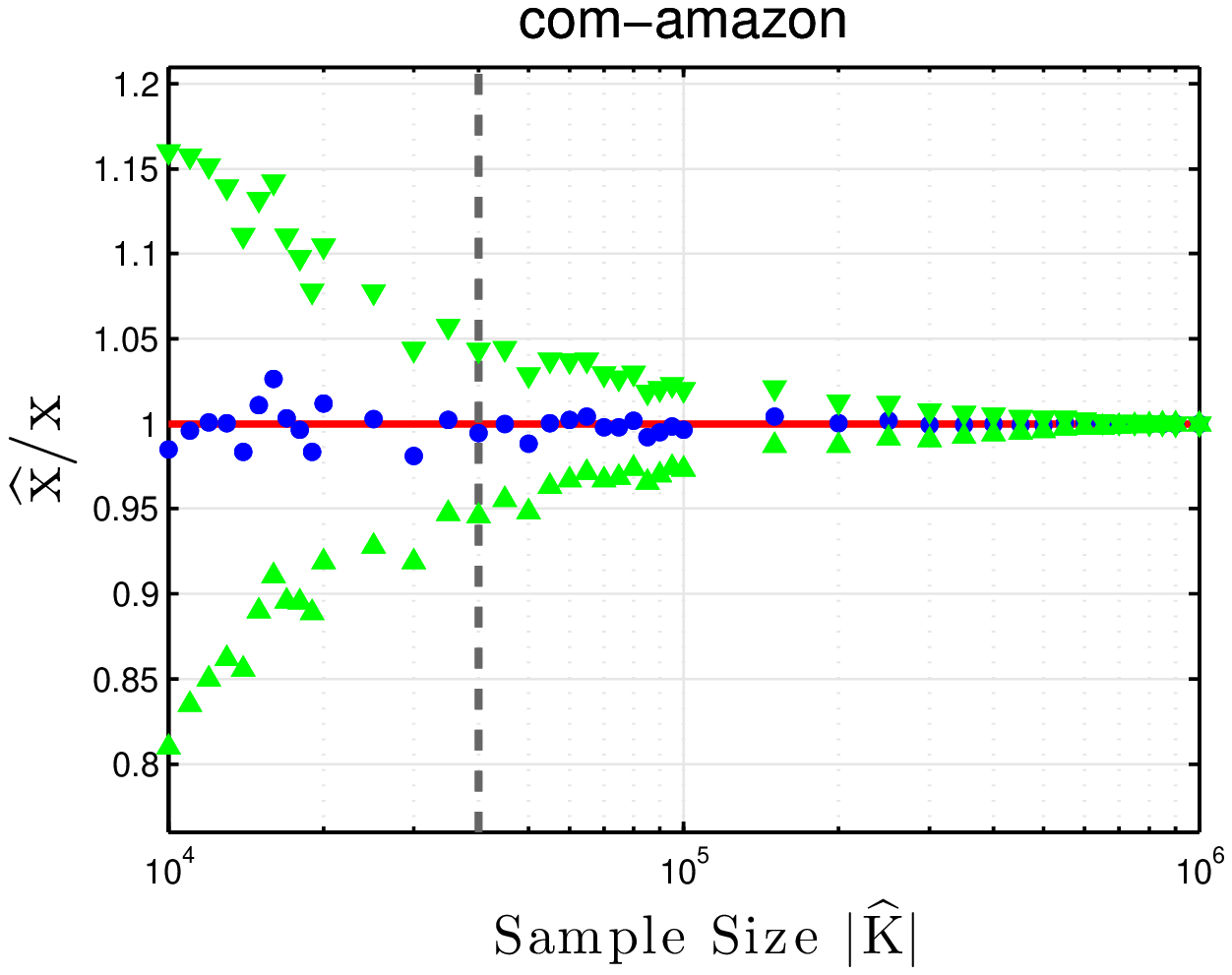}}
\hspace*{\figConfhspace}\subfigure	
{\includegraphics[width=\figConfSZ]{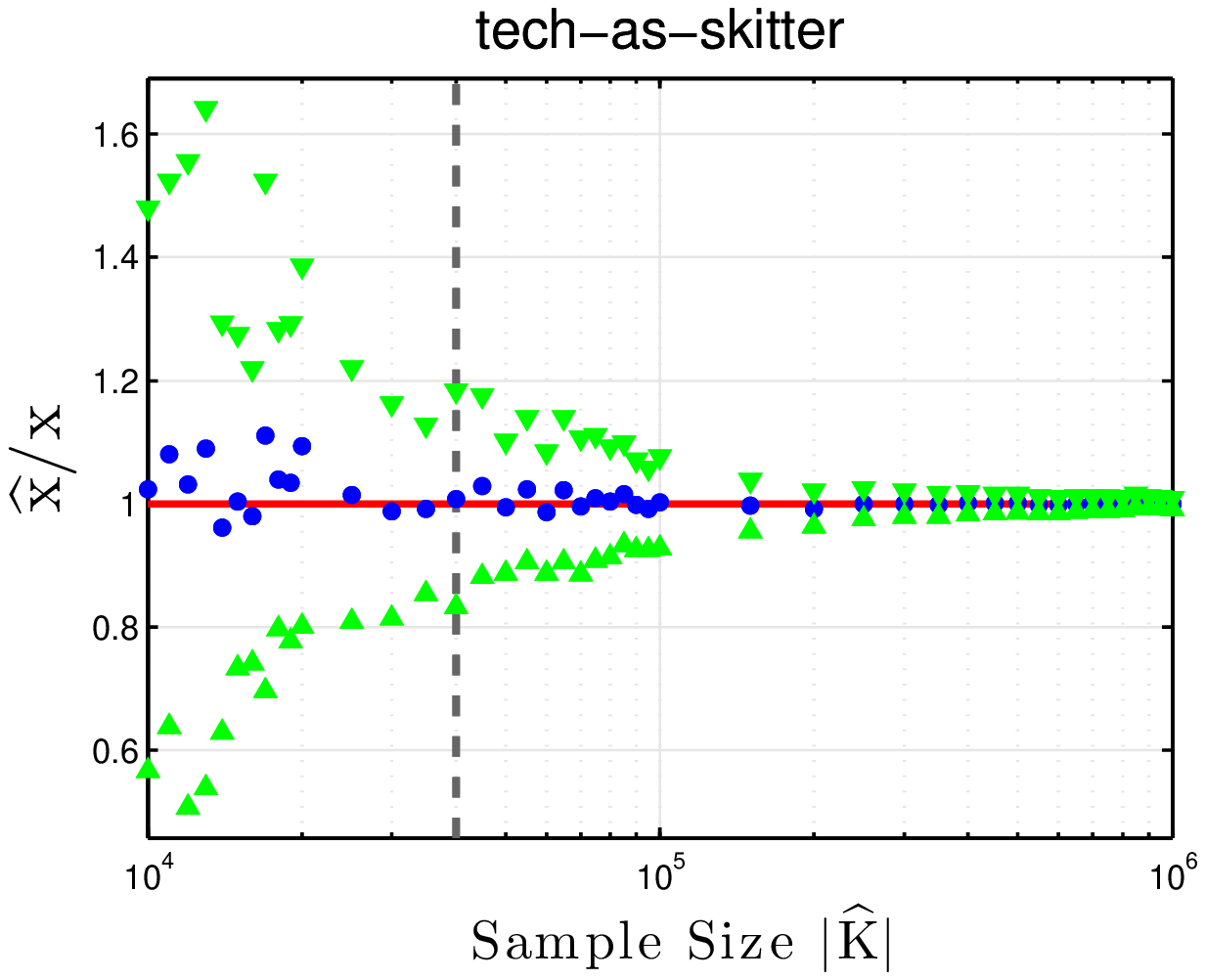}}
\hspace*{\figConfhspace}\subfigure		
{\includegraphics[width=\figConfSZ]{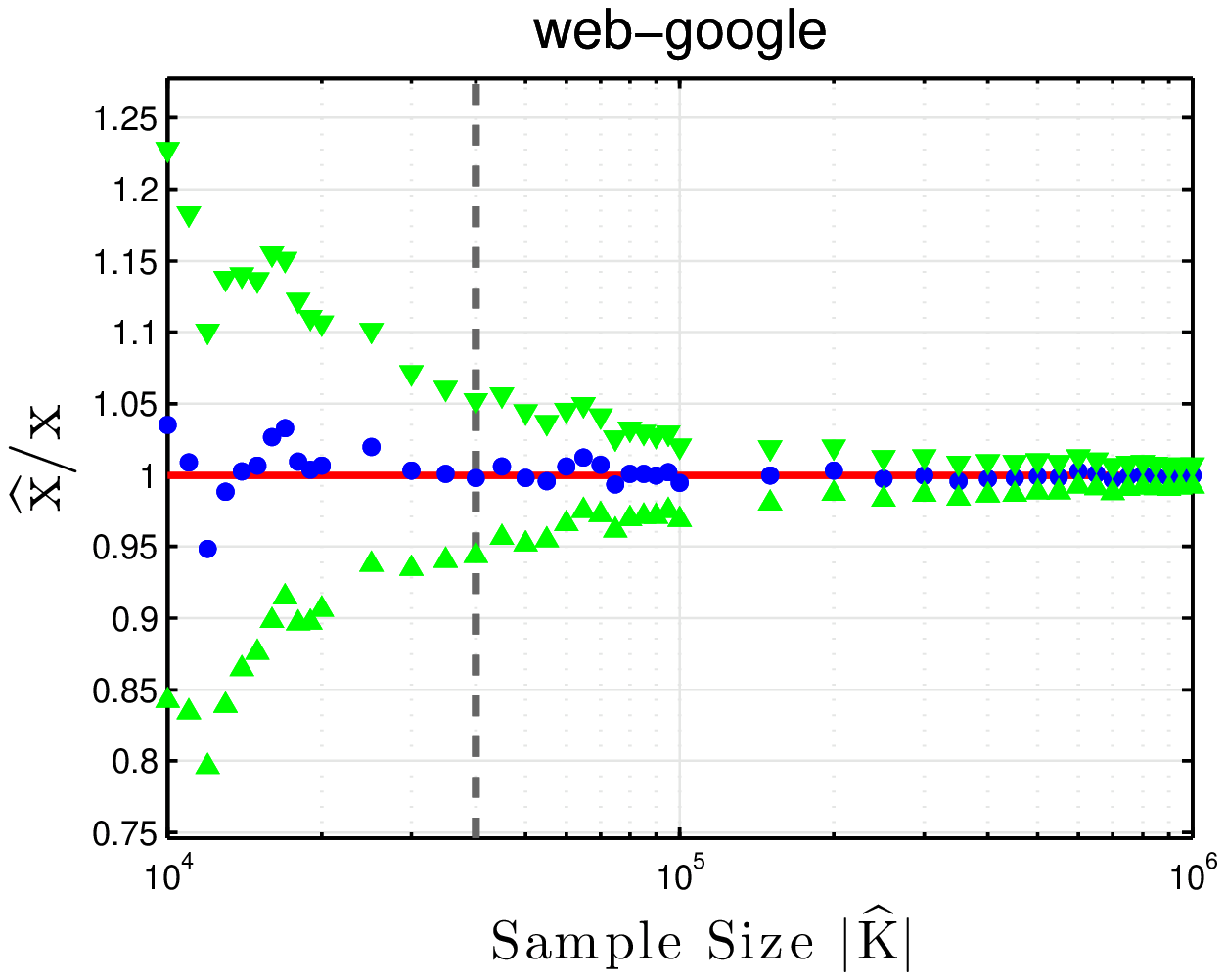}}

\vspace{-2.5mm}
\caption{Confidence bounds for Graph Priority Sampling with instream estimation of triangle counts. 
We used graphs from a variety of domains and types. The properties of the sampling distribution and convergence of the estimates are investigated as the sample size increases. The circle (\textcolor{blue}{$\newmoon$}) represents $\nicefrac{\widehat{X}}{X}$ (y-axis) whereas \textcolor{green}{$\blacktriangle$} and \textcolor{green}{$\blacktriangledown$} are $\nicefrac{LB}{X}$ and $\nicefrac{UB}{X}$, respectively. Dashed vertical line (grey) refers to the sample at $40$K edges. Notably, the proposed framework has excellent accuracy even at this small sample size.
}
\label{fig:gps-triangle-confidence-bounds}
\vspace{-3mm}
\end{figure*}

{
\setlength{\tabcolsep}{3.06pt}
\begin{table}[t!]
\vspace{-7.mm}
\parbox[c]{1.0\linewidth}{
\begin{center}
\scriptsize
\caption{
Baseline Comparison at sample size $\approx$100K}
\label{tab:baseline_comp_all}
\scalebox{1.0}{
\begin{tabular}{lcccc}
\toprule
& \textsc{Nsamp} & \textsc{TRIEST} & \textsc{MASCOT}& \textsc{GPS Post} \\
\midrule
& \multicolumn{4}{c}{Absolute Relative Error (ARE)} \\ 
\cmidrule{2-5} 
\dataName{cit-Patents} &  $0.192$ &  $0.401$ & $0.65$ & $0.008$ \\
\dataName{higgs-soc-net} & $0.079$ & $0.174$ & $0.209$ & $0.011$ \\ 
\dataName{infra-roadNet-CA} & $0.165$  & $0.301$ & $0.39$ & $0.013$\\
\midrule
& \multicolumn{4}{c}{Average Time ($\mu_s$/edge)} \\ 
\cmidrule{2-5}
\dataName{cit-Patents} &  $34.2$ &  $3.01$ & $2.02$ & $0.63$ \\
\dataName{higgs-soc-net} & $26.08$ & $4.40$ & $2.02$ & $11.74$ \\ 
\dataName{infra-roadNet-CA} & $28.72$  & $2.81$ & $2.05$ & $0.831$\\
\bottomrule
\end{tabular}}
\end{center}}
\vspace{-7.mm}
\end{table}
}

\parab{Baseline Study.} The state-of-the-art algorithms for triangle count estimation in adjacency graph streams are due to the neighborhood sampling (\textsc{Nsamp}) in~\cite{pavan2013counting} and the triangle sampling (\textsc{TRIEST}) in~\cite{de2016tri}. We discuss their performance in turn compared with \gps\ post stream estimation. We also compare with \textsc{MASCOT}~\cite{lim2015mascot}. Table~\ref{tab:baseline_comp_all} summarizes the results of the comparison. Our implementation of the \textsc{Nsamp}~\cite{pavan2013counting} algorithm follows the description in the paper, which achieves a near-linear total time if and only if running in bulk-processing. Otherwise the algorithm is too slow and not practical even for medium size graphs with a total time of $\mathcal{O}(|K| r)$.
Overall, \gps\ post stream estimation achieves 98\%--99\% accuracy, while \textsc{Nsamp} achieves only 80\%--84\% accuracy for most graphs and 92\% accuracy for higgs-soc-net graph. Our implementation of the \textsc{TRIEST}~\cite{de2016tri} algorithm follows the main approach in the paper. \textsc{TRIEST} was unable to produce a reasonable estimate showing only 60\%--82\% accuracy. Similarly, \textsc{MASCOT} achieves only 35\%--79\% accuracy. Thus, \gps\ post stream estimation outperforms the three baseline methods. Table~\ref{tab:baseline_comp_all} also shows the average update time per edge (in microseconds). We note that \gps\ post stream estimation achieves an average update time that is 35x--56x faster than \textsc{Nsamp} with bulk-processing (for cit-Patents and infra-roadNet-CA graphs and at least 2x faster for higgs-soc-net). \textsc{TRIEST} and \textsc{MASCOT} use an average of 3 and 2 microseconds/edge respectively. 
Note that we have also compared to other methods in~\cite{jha2013space} and~\cite{buriol2006counting} (results omitted for brevity). Even though the method in ~\cite{buriol2006counting} is fast, it fails to find a triangle most of the time, producing low quality estimates (mostly zero estimates). On the other hand, the method of~\cite{jha2013space} is too slow for extensive experiments with $\mathcal{O}(m)$ update complexity per edge (where $m$ is the reservoir size). \gps\ post stream estimation achieves at least 10x accuracy improvement compared to their method. For this comparison, we first run \textsc{MASCOT} with approximately 100K edges, then we observe the actual sample size used by \textsc{MASCOT} and run all other methods with the observed sample size.

\begin{figure}
\vspace{-4mm}
\centering
\subfigure	
{\includegraphics[width=0.49\linewidth]{{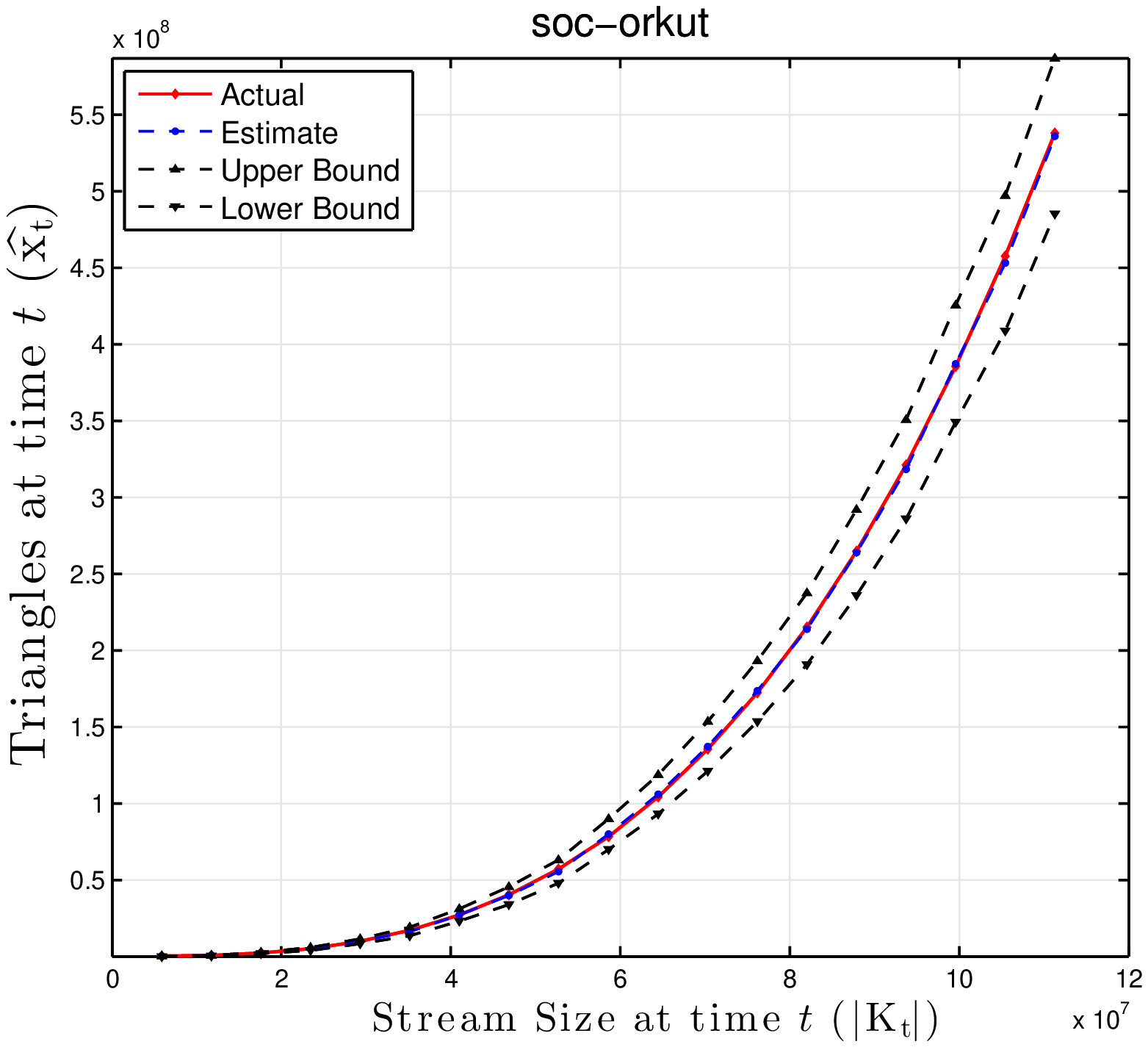}}}
\hfill
\subfigure	
{\includegraphics[width=0.45\linewidth]{{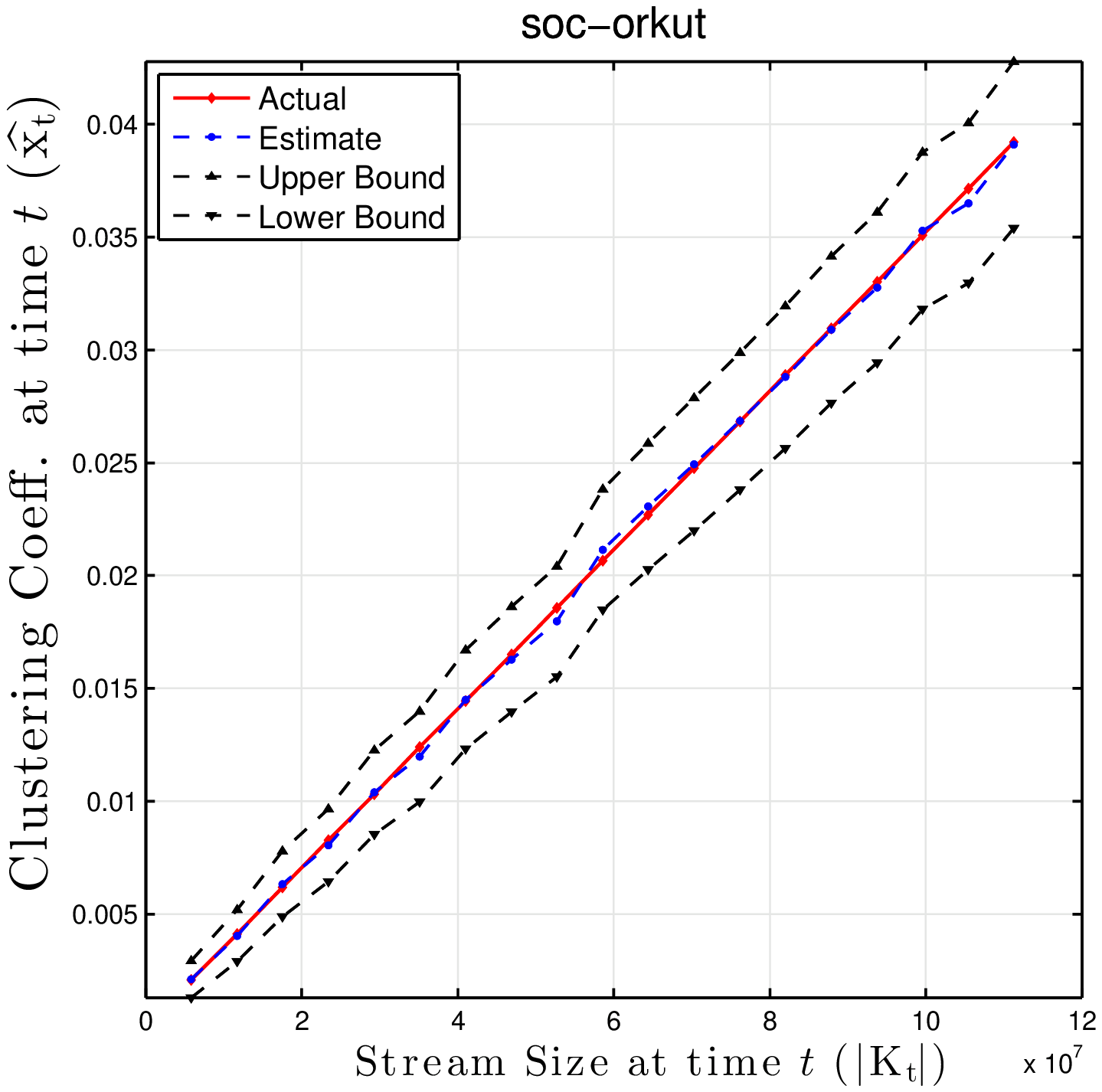}}}
\vspace*{-3mm}

\subfigure
{\includegraphics[width=0.49\linewidth]{{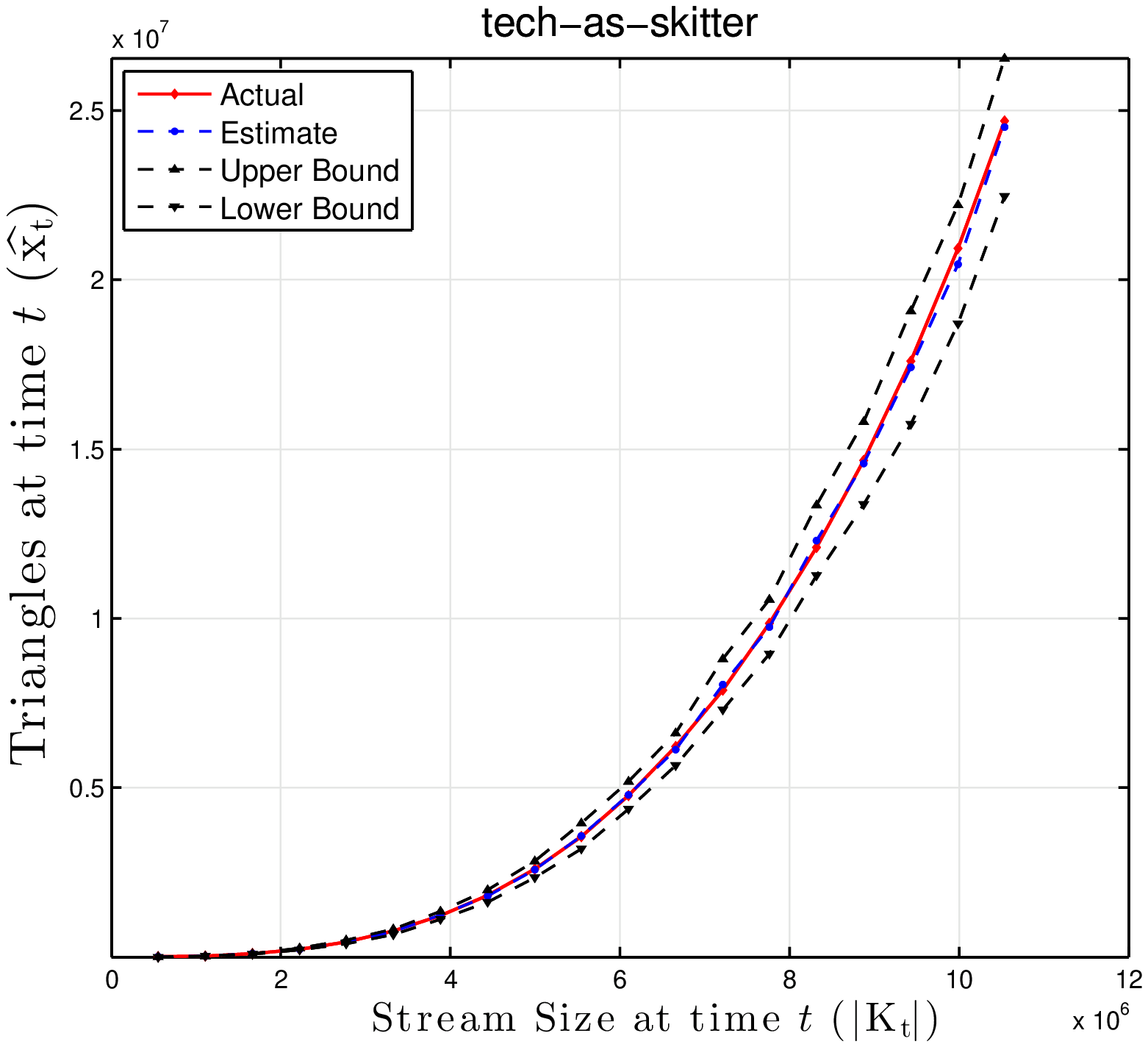}}}
\hspace*{0.3mm}
\hfill
\hspace*{0.16mm}
\subfigure	
{\includegraphics[width=0.45\linewidth]{{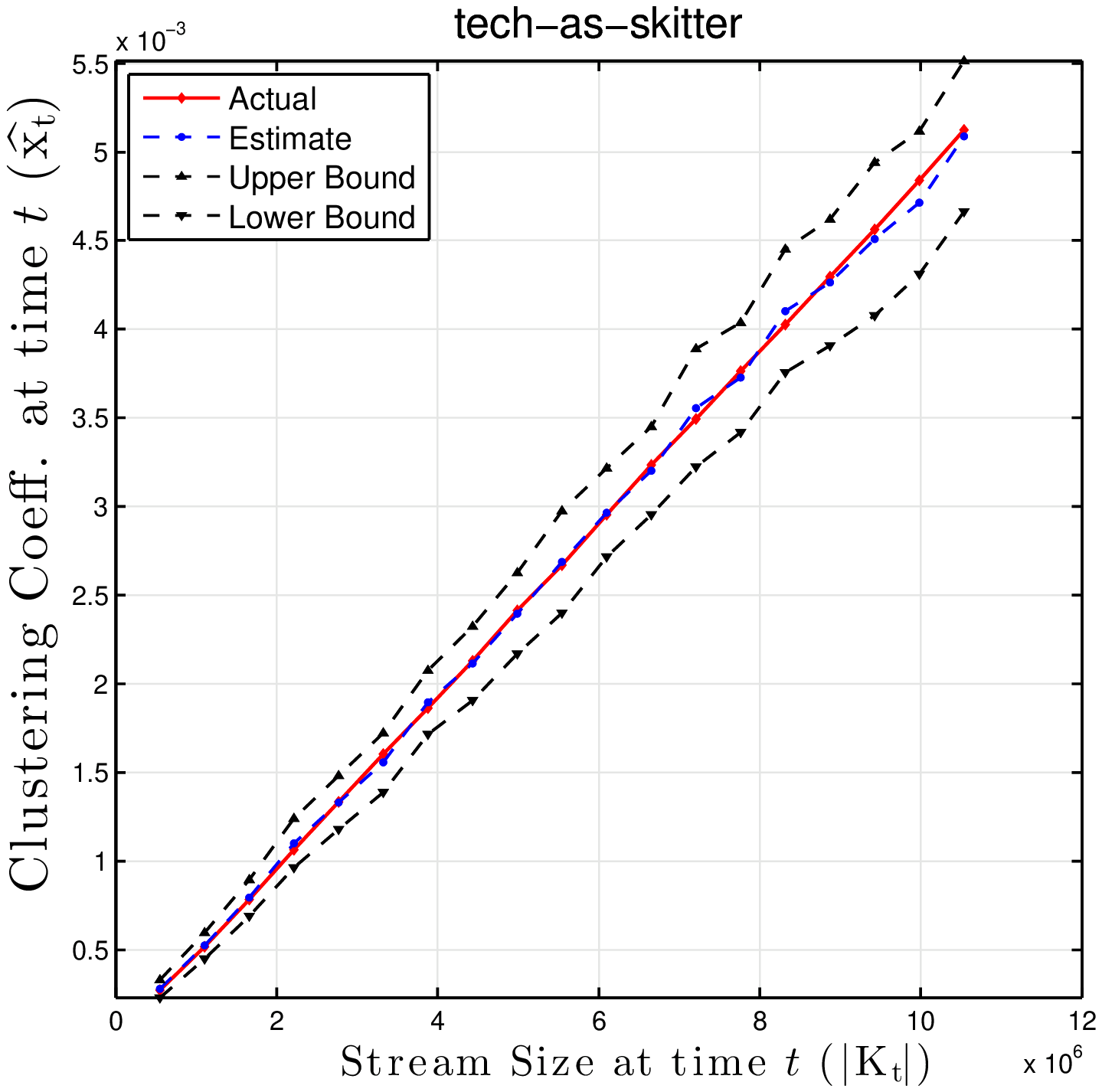}}}

\vspace{-2.0mm}
\caption{Graph Priority Sampling with in-stream estimation versus time. Results for social and tech networks at sample size 80K edges for triangle counts and global clustering with $95\%$ confidence lower and upper bounds. Notably, the proposed framework accurately estimates the statistics while the stream is progressing.}
\label{fig:gps-est-vs-time}
\vspace{-5mm}
\end{figure}

\parab{Unbiased Estimation Vs. Time.} We now track the estimates as the graph stream progresses one edge at a time, starting from an empty graph. Figure~\ref{fig:gps-est-vs-time} shows \gps\ estimates for triangle counts and clustering coefficient as the stream is evolving overtime. Notably, the estimates are indistinguishable from the actual values. Figure~\ref{fig:gps-est-vs-time} also shows the 95\% confidence upper and lower bounds. These results are for a sample of 80K edges (a small fraction $\leq 1\%$ of the size of soc-orkut and tech-as-skitter graphs) using \gps\ in-stream

Given the slow execution of \textsc{Nsamp}, we compare against \textsc{TRIEST} and its improved estimation procedure (\textsc{TRIEST-Impr}). Note that \textsc{TRIEST} and \textsc{TRIEST-Impr} are both using the same sample and random seeds. The two approaches are based on reservoir sampling~\cite{Vitter:85}. We used graphs from a variety of domains and types for this comparison, and all methods are using the same sample size. We measure the error using the Mean Absolute Relative Error (MARE) $\frac{1}{T} \sum_{t=1}^{T} \frac{|\hat X_t - X_t|}{X_t}$, where $T$ is the number of time-stamps. We also report the maximum error $\max_{t=1}^{T} \frac{|\hat X_t - X_t|}{X_t}$. Table~\ref{tab:baseline_time_comp} summarizes the comparison results. For all graphs, \gps\ with in-stream estimation outperforms both \textsc{TRIEST} and its improved estimation procedure \textsc{TRIEST-Impr}. We note that indeed \textsc{TRIEST-Impr} significantly improves the quality of the estimation of \textsc{TRIEST}. However, in comparison with \gps\, we observe that \gps\ with post stream estimation is orders of magnitude better than \textsc{TRIEST}, which shows that the quality of the sample collected by \gps\ is much better than \textsc{TRIEST} (regardless the estimation procedure used, whether it is post or in-stream estimation). 

\parab{Scalability and Runtime.} Our Algorithm~\ref{alg:alg_gpsest} for post steam estimation uses a scalable parallel approach from~\cite{ahmed2015efficient,pgd-kais} with strong scaling properties; we omit the scalability results for brevity.

{
\setlength{\tabcolsep}{2.40pt}
\begin{table}[t!]
\vspace{-6mm}
\parbox[c]{1.0\linewidth}{
\begin{center}
\scriptsize
\caption{
Mean absolute relative error for estimates of triangle counts vs. time (sample size = 80K).
}
\vspace{1mm}
\label{tab:baseline_time_comp}
\scalebox{1.0}{
\begin{tabular}{llcc}
\toprule
\textsf{graph} & \textsf{Algorithm} &  \textsf{Max. ARE} & \textsf{MARE} \\
\midrule
\multirow{4}{*}{\dataName{ca-hollywood-2009}} & \textsc{TRIEST} & $0.492$ & $0.211$ \\
& \textsc{TRIEST-Impr} & $0.066$ & $0.018$ \\
& \textsc{GPS Post} & $0.049$ & $0.020$ \\
& \textsc{GPS in-stream} & $0.016$ & $0.003$ \\
\midrule
\multirow{4}{*}{\dataName{tech-as-skitter}} & \textsc{TRIEST} & $0.628$ & $0.249$ \\
& \textsc{TRIEST-Impr} & $0.134$ & $0.048$ \\
& \textsc{GPS Post} & $0.087$ & $0.035$ \\
& \textsc{GPS in-stream} & $0.032$ & $0.014$ \\
\midrule
\multirow{4}{*}{\dataName{infra-roadNet-CA}} & \textsc{TRIEST} & $0.98$ & $0.47$ \\
& \textsc{TRIEST-Impr} & $ 0.33$ & $0.09$ \\
& \textsc{GPS Post} & $0.15$ & $0.05$ \\
& \textsc{GPS in-stream} & \emph{$0.058$} & $0.02$ \\
\midrule
\multirow{4}{*}{\dataName{soc-youtube-snap}} & \textsc{TRIEST} & $0.362$ & $0.119$ \\
& \textsc{TRIEST-Impr} & $0.049$ & $0.016$ \\
& \textsc{GPS Post} & $0.022$ & $0.009$ \\
& \textsc{GPS in-stream} & \emph{$0.020$} & $0.008$ \\
\bottomrule
\end{tabular}}
\end{center}}
\vspace{-7.mm}
\end{table}
}

\section{Related Work}
\label{sec-related}

The research related to this paper can be grouped into: (a) Sampling from graph streams, (b) Sampling/Estimation of subgraphs.

\parab{Sampling from graph streams.}
Recently, there has been considerable interest in designing algorithms for mining massive and dynamic graphs from data streams, motivated by applications.
Many were based on 
sampling and approximation techniques. An early such work 
\cite{raghavan1999computing} concerned problems of following paths and connectivity in directed graphs. Much of the earlier work on graph streams focused on graph problems in the semi-streaming model~\cite{muthu,feigenbaum2005graph}, where the algorithm is allowed to use $\mathcal{O}(n\, \polylog\, n)$ space to solve graph problems that are provably intractable in sub-linear space. More recent work focused on graph mining problems such as finding common neighbors~\cite{buchsbaum2003finding}, estimation of pagerank~\cite{atish}, clustering and outlier detection~\cite{aggarwal2011outlier}, characterizing degree sequences in multigraph streams~\cite{cormode2005space}, link prediction~\cite{Zhaolinkpred}, community detection~\cite{zakrzewska2016tracking}, among others~\cite{mcgregor2009graph,rossi2014roles,ahmed2010time,ahmed2012network}. See~\cite{AhmedTKDD} for further details.

\parab{Sampling and Estimation of Subgraphs.} 
Subgraph counting (in particular triangle counting) has gained significant attention due to its applications in social, information, and web networks. 
Early work in this direction \cite{buriol2006counting} provides a space-bounded algorithm for the estimation of triangle counts and clustering coefficient in the incidence graph stream model. However, it has been shown that these guarantees will no longer hold in the case of the adjacency stream model, where the edges arrive in arbitrary 
order.
A single pass streaming algorithm 
incurring $\mathcal{O}(m \Delta / T)$-space, where $\Delta$ is the maximum degree is proposed in \cite{pavan2013counting}. However, this algorithm requires both large storage and update overhead to provide accurate results. 
For example, their algorithm needs at least $128$ estimators (\ie, storing more than $128$K edges) and uses large batch sizes (\eg, a million edges) to obtain accurate/efficient results. 
A single-pass $\mathcal{O}(m / \sqrt{T})$-space streaming algorithm was proposed in
\cite{jha2013space} specifically for transitivity estimation with arbitrary additive error guarantees. This algorithm maintains two reservoirs, the first to select a uniform sample of edges from the stream, and the second to select a uniform sample of wedges created by the edge reservoir. 

Other approaches focused on maintaining a set of edges sampled randomly from the graph stream. \emph{graph sample-and-hold} \cite{Ahmed-gSH} is a framework for unbiased an accurate estimation of subgraph counts (\eg, edges, triangles, wedges). 
A similar approach was recently proposed for local (node/edge) triangle count estimation in graph streams \cite{lim2015mascot}. Other methods extend reservoir sampling to graph streams. 
For example, reservoir sampling has been used for detecting outliers in graph streams~\cite{aggarwal2011outlier}, estimating the distribution of various graph properties (\eg, path length, clustering)~\cite{AhmedTKDD}, and estimating triangle counts in dynamic graph streams with insertions and deletions \cite{de2016tri}.
\section{Conclusion}
\label{sec:conclude}

In this paper, we presented \emph{graph priority sampling}, a new framework for order-based reservoir sampling from massive graph streams. \gps\ provides a general way to weight edge sampling according to auxiliary variables to estimate various graph properties. We showed how edge sampling weights can be chosen so as to minimize the estimation variance of counts of subgraphs, such as triangles and wedges. Unlike previous graph sampling algorithms, \gps\ differentiates between the functions of sampling and estimation. We proposed two estimation approaches: (1) Post-stream estimation, to allow \gps\ to construct a reference sample of edges to support retrospective graph queries, and (2) In-stream estimation, to allow \gps\ to obtain lower variance estimates by incrementally updating the count estimates during stream processing. We provided a novel Martingale formulation for subgraph count estimation. We performed a large-scale experimental analysis. The results show that \gps\ achieves high accuracy with $< 1\%$ error on large real-world graphs from a variety of domains and types, while storing a small fraction of the graph and average update times of a few microseconds per edge. In future work, we aim to extend the proposed approach to adaptive-weight sampling schemes and its applications in massive streaming analytics.

\section{Proofs of Theorems}\label{sec-proofs}
{\vspace*{-2.8mm}
\begin{lemma}\label{lem:meas}
For each $t$, the events $\left\{\{j\in K_t\}: j\le t\right\}$ are measurable w.r.t $\cF_{i,t}$ .
\end{lemma}
}
\begin{proof}[of Lemma~\ref{lem:meas}] The proof is by induction. It is trivially true fot
  $t=i$. Assume true for $t-1$, then membership of $ K_{t-1}$ is
  $\cF_{i,t-1}$ measurable, and hence so is membership of
  $ K'_{t-1}$. Selection of $i$ is clearly
  $\cF_{i,t}$-measurable, and if $i$ is selected, the remaining
  selections are $\cF\up 0_{i,t}$-measurable since then
  $z_{\{ij\},t}=z_{j,t}$.
\end{proof}
\begin{proof}[of Theorem~\ref{thm:single}]
Lemma~\ref{lem:meas} established measurability.\\
For $t\ge i$, 
since
$R_{i,t}$ is $\cZ_{i,t}$-measurable,
conditioning first on $z_{i,t}$:
\be
\E[\hat S_{i,t} | z_{i,t},\cF_{i,t-1}] 
=\frac{1}{R_{i,t}}\E[I(B_i(z^*_{i,t}))|z_{i,t},\cF_{i,t-1}]
\label{eq:ce1}\ee
Since $B_i(z^*_{i,t})=B_i(z_{i,t})\cap B_i(z^*_{i,t-1})$ and $I(B_i(z^*_{t-1}))$ is $\cF_{i,t-1}$-measurable, then for any
event on which the conditional expectation (\ref{eq:ce1}) is positive,
we have
\bea 
&& \kern -25pt \E[I(B_i(z^*_{i,t}))|z_{i,t},\cF_{i,t-1}] 
=\Pr[B_i(z_{i,t})|B_i(z^*_{i,t-1}),\cZ_{i,t},\cF\up 0_{i,t-1}]
\nonumber \\
&=&{\Pr[B_i(z^*_{i,t})|\cZ_{i,t},w_i]}/
{\Pr[B_i(z^*_{i,t-1})|\cZ_{i,t},w_i]}\nonumber\\
&=&{\Pr[w_i/u_i > z^*_{i,t}|\cZ_{i,t},w_i]}/
{\Pr[w_i/u_i > z^*_{i,t-1}|\cZ_{i,t},w_i]}\nonumber\\
&=&R_{i,t}/R_{i,t-1}
\eea
where we have used the fact that once we have conditioned on
$B_i(z^*_{i,t-1})$ and $\cZ_{i,t}$, then $\cF\up0_{i,t-1}$ conditions
only through the dependence of $w_i$ on the
sample set $ K_{i-1}$.
Thus we have established that $\E[\hat S_{i,t}|
z_{i,t},\cF_{i,t-1}]=\hat S_{i,t-1}$ regardless of $z_{i,t}$, and
hence $\E[\hat S_{i,t}|\cF_{i,t-1}]=\hat S_{i,t-1}$.
\end{proof}

\begin{proof}[of Theorem~\ref{thm:snap:unb}] (ii) follows from (i) by
  linearity of expectation. For (i), observe that
$\hat S^{T_j}_{j,t}= S^{T_j}_{j,t-1} + I(T\ge t)(\hat S_{j,t}-\hat
S_{j,t-1})$; one checks that this reproduces $\hat S_{j,\min\{t,T_j\}}$. Thus
\be\nonumber
\hat S_{J,t}^{T}=\prod_{j\in J} \hat S_{j,t-1}^{T_j}+\sum_{L\subsetneq
  J} \prod_{\ell\in L} S^{T_\ell}_{j,\ell-1}\kern -5pt \prod_{j\in J\setminus L}
I(T_j\ge t) (\hat S_{j,t}-\hat
S_{j,t-1})
\ee
Observe that $I(T_j\ge t)=1-I(T_j\le t-1)$ is in fact
$\cF_{J,t-1}$-measurable. Hence taking expectations w.r.t.
$\cF_{J,t-1}$ then the product form from Theorem~\ref{thm:prod} tells
us that for any $L\subsetneq J$
\bea
\E[ \prod_{j\in J\setminus L}
I(T_j\ge t) (\hat S_{j,t}-\hat
S_{j,t-1})|\cF_{J,t-1}] \kern -150pt&&
\nonumber \\ 
&=&\prod_{j\in J\setminus L}
I(T_j\ge t) \E[\hat S_{j,t}-\hat
S_{j,t-1}|\cF_{J,t-1}]=0
\eea
and hence
$
\E[\hat S_{J,t}^{T} |\cF_{J,t-1}] = \prod_{j\in J} \hat
S_{j,t-1}^{T_j} = \hat S_{J,t-1}^{T}
$
\end{proof}

\begin{proof}[of Theorem~\ref{thm:snap:cov}] (i) 
\bea \cov( \hat S^{T\up 1}_{J_1,t} ,\hat S^{T\up
    2}_{J_2,t})&=&\E[\hat S^{T\up
    1}_{J_1,t} \hat S^{T\up 2}_{J_2,t}]-\E[\hat S^{T\up
    1}_{J_1,t}]\E[\hat S^{T\up 2}_{J_2,t}]\nonumber\\
&=& \E[\hat S^{T\up
    1}_{J_1,t} S^{T\up 2}_{J_2,t}]-1\eea 
Hence the results follows because
\be
E[\hat S^{T\up 1}_{J_1\setminus J_2,t}\hat S^{T\up
  2}_{J_2\setminus J_1,t}\hat S^{T\up 1\vee T\up 2}_{J_1\cap J_2,t}]=1
\ee
from Theorem~\ref{thm:snap:unb} since $J_1\setminus J_2$,
$J_2\setminus J_1$ and $J_1\cap J_2$ are disjoint.

(ii)$\hat C^{T\up 1,T\up 2}_{J_1,J_2,t}=
\hat S^{T\up 1}_{J_1\setminus J_2,t} \hat S^{T\up
  2}_{J_2\setminus J_1,t} (\hat S^{T\up 1}_{J_1\cap J_2}\hat
  S^{T\up 2}_{J_1\cap J_2}   -  \hat S^{T\up 1\vee T\up 2}_{J_1\cap J_2,t})$.
which is nonnegative since each $j\in J_1\cap J_2$ brings a factor of
the form $1/(p\up 1 p\up 2)$ to $\hat S^{T\up 1}_{J_1\cap J_2}\hat
  S^{T\up 2}_{J_1\cap J_2}$, where $p\up i=p_{j,\max\{t,T\up
      i_j\}}$. This exceeds the matching term in $\hat S^{T\up 1\vee T\up
    2}_{J_1\cap J_2,t}$, i.e., $1/\min\{p_1,p_2\}$.

(iii) Follows from (i) upon setting $J_1=J_2=J$.

(iv)  The equivalence clearly applies to the first monomial in
(\ref{eq:cov:est:snap}). For the second monomial, note that 
$(\hat S^{T\up 1}_{J_1,t}=0)\wedge(\hat S^{T\up 2}_{J_2,t}=0)$ if
and only if $\hat S^{T\up 1}_{j,t}=0$ for some $j\in J_1$ or $\hat S^{T\up
  2}_{j,t}=0$ for some $j\in J_2$. If this condition holds for some
$j\in J_1\Delta J_2$ we are done. Otherwise, we require $\hat S^{T\up
  i_j}_{j,t}=0$ for some $j\in J_1\cap J_2$ and some $i\in\{1,2\}$. But
for $j\in J_1\cap J_2$,
$\hat S^{T\up i_j}_{j,t}=\hat S_{j,\min\{T\up i_j,t\}}=0$ means that
$j$ has not survived until $\min\{T\up i_j,t\}$ and hence it also not
present at the later or equal time  $\min\{\max\{T\up 1_jT\up
2_j\},t\}$. Hence $\hat S^{\max\{T\up 1_j,T\up 2_j\}}_{j,t}=0$ and we
are done
\end{proof}

\begin{proof}[of Theorem~\ref{thm:tri:instream}]
$\hat S^{T_{k_3}}_{\{k_1,k_2\},t}>0$ only if
$(k_1,k_2,k_3)\in\Delta_t$, 
and by Theorem~\ref{thm:snap:unb} has unit expectation.
\end{proof}

\begin{proof}[of Theorem~\ref{thm:tri:instream:var}]
Distributing the
  covariance over the \\ 
sum $\tilde N_t(\vartriangle)$ in
  Theorem~\ref{thm:tri:instream}, 
$\var(\tilde N_t(\vartriangle))$ has unbiased estimator 

\be\label{eq:tri:instream} \kern -5pt
\sum_{(k_1,k_2,k_3)\in \vartriangle_t}\kern -17pt 
\hat S_{\{k_1,k_2\},t}^{T_{k_3}}(\hat S_{\{k_1,k_2\},t}^{T_{k_3}}
-1)+2 \kern -15pt  \sum_{ {(k_1,k_2,k_3)\prec \atop (k'_1,k'_2,k'_3)}} \kern -15pt
C_{\{k_1,k_2\},\{k_1',k_2'\},t}^{T_{k_3},T_{k'_3}}
\ee

Where $\prec$ denotes $k_3<k'_3$ in arrival order.
Each variance term is of the form $1/q(1/q-1)$ for
$q=p_{k_1,T_3}p_{k_2,T_3}$. Each covariance term is zero unless
$\{k_1,k_2\}\cap \{k'_1,k'_2\}\ne\emptyset$. (The sets cannot be equal for then
$k_3=k'_3$ if both form triangles). The stated form the follows by
rewriting the sum of covariance terms in (\ref{eq:tri:instream})
as a sum over edges $k\in\tilde K_t$ of the covariances of all
pairs snapshots that contain $k$ as a sampled edge.
\end{proof}

\bibliographystyle{abbrv}
\bibliography{paper,cycle,ahmed,bibmaster}
\end{document}